\documentclass[%
preprint,
preprintnumbers,
 amsmath,amssymb,
 prfluids,
]{revtex4-2}

\usepackage{subfigmat}
\usepackage{etoolbox}
\patchcmd{\subfigmatrix}{\hfill}{\hspace{0.01cm}}{}{}
\usepackage{graphicx}

\usepackage{soul}
\usepackage{xargs}
\usepackage[colorinlistoftodos,prependcaption,textsize=tiny]{todonotes}

\renewcommand{\todo}[1]{}

\newif\ifproofread

\newcommand{\new}[1]{%
\ifproofread
\textcolor{red}{#1}%
\else
#1%
\fi
}

\begin{document}
\proofreadfalse  

\title{\new{The} role of entropic-instabilities in laminar-turbulent transition on a blunted flat plate}

\author{Hemanth Goparaju}
 \email{goparaju.3@osu.edu}
 \author{Datta V Gaitonde}
 \email{gaitonde.3@osu.edu}
\affiliation{%
 Department of Mechanical and Aerospace Engineering \\
 The Ohio State University, Columbus, OH, 43210 USA\\
}%


\begin{abstract}
The effects of entropic-instabilities on the laminar-turbulent transition dynamics of a blunted flat plate at Mach~$4$ are numerically investigated through linear and nonlinear approaches.
Linear wavepacket analysis reveals amplifying oblique first-modes as well as planar and oblique entropy-layer disturbances.
The receptivity of entropic-instabilities is found to be largest for actuation seeded in the entropy-layer; the corresponding linear evolution is characterized by an intensification in the wall-normal plane, coupled with streamwise tilting. 
\new{The transition process that arises from entropy-layer wave interactions via the oblique breakdown mechanism is examined in detail.}
Oblique entropy-layer disturbances interact non-linearly with each other and induce streamwise streaks in the boundary-layer. 
These undergo further destabilization downstream, through a combination of lift-up and Orr-like mechanisms, leading to turbulence onset.
\new{Sinuous subharmonic oscillations induced by the entropic-disturbances are the dominant streak instabilities. 
Slanted \textit{hook}-shaped structures are observed in the temperature perturbations, and are attributed to an Orr-like mechanism in the entropy-layer.
Further, temperature perturbations amplify on the crests of the low-speed streaks in the entropy-layer and generate disturbances through triadic interactions. 
These disturbances aid in the transfer of perturbation energy into the boundary-layer, eventually leading to spanwise homogenization and near-wall streak generation.}
Towards the end of the breakdown region, spectral broadening accompanied by the appearance of an inertial sub-range in the boundary-layer indicates the approach of the flow towards turbulence.
\new{The prominence of entropy-layer disturbances in streak generation and destabilization suggests oblique breakdown as a potentially important transition mechanism in high-speed blunt-nosed configurations.}
\end{abstract}

\maketitle
\section{Introduction}
\label{sec:intro}
Efficient hypersonic vehicle design is predicated on accurate prediction of the location of boundary-layer laminar–turbulent transition, which substantially influences aerodynamic heating and skin friction.
This requires a comprehensive understanding of the underlying fundamental transition mechanisms.
The leading edges of practical aerospace vehicles are blunted to mitigate peak heat transfer rates.
Transition mechanisms on such geometries are not as yet fully understood, and are the focus of this work.

\begin{figure}
    \centering
    \includegraphics[width=0.8\textwidth]{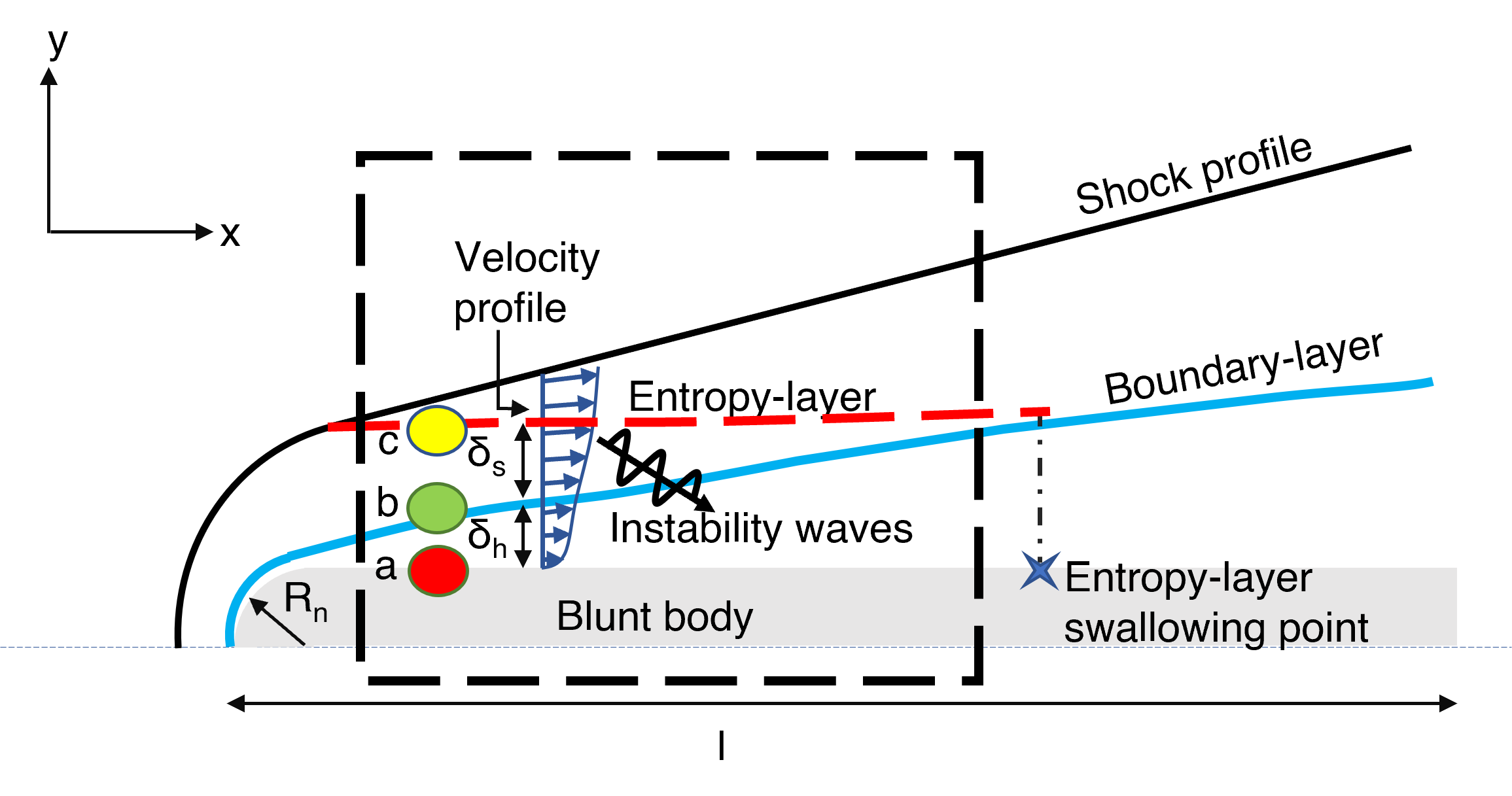}
    \caption{Schematic of the flow features on a leading edge blunted plate.
    The dashed box indicates the sub-domain extracted for high-fidelity numerical simulations.
    Perturbations imposed in the current study are at the $(a)$~Wall $(b)$~Boundary-layer edge $(c)$~Entropy-layer edge.
    Sketch adapted from Zhong and Ma~\cite{zhong2006boundary}.}
    \label{fig:schematic}
\end{figure}

In blunted nose configurations, a curved detached shock generates an  entropy gradient near the leading-edge, resulting in an \textit{entropy-layer}.
This is schematically depicted in fig.~\ref{fig:schematic}, which also displays a developing boundary-layer on the plate, and other features of interest which are discussed below.
On a long enough body, the entropy-layer is eventually absorbed by the growing boundary-layer; the corresponding streamwise extent is the \textit{swallowing length}.

Experimental investigations on hypersonic blunt plates~\cite{lysenko1990influence,borovoy2021laminar} and cones~\cite{stetson1983nosetip,jewell2018transition} have observed that as the  nose radius is increased, the transition location first moves downstream and then upstream (\textit{transition reversal}).  
Based on the nose radius, Stetson~\cite{stetson1983nosetip} classified transition on blunt cones into three categories, \textit{viz.}, small, medium, and large bluntness regimes.
In the small bluntness regime, the transition location moves downstream with increasing nose radii; the dominant instability mechanism is then first- or second-mode~\cite{mack1984boundary} induced.
In the medium bluntness regime, the transition location also moves downstream and the boundary-layer modes are unstable with increasing radii; however, their growth-rate is too low to cause transition as observed in experiments.
The large bluntness regime has no unstable boundary-layer modes but transitions upstream with increasing nose radii.
Stetson~\cite{stetson1983nosetip} also attributed the transition in the medium and large-bluntness regimes to the amplification of disturbances in the entropy-layer. 

Transitional flow features on blunt cones were further investigated experimentally by Grossir et al.~\cite{grossir2014hypersonic} at Mach~$11.9$ and Kennedy et al.~\cite{kennedy2019visualizations} at Mach~$6$, with laser-induced-fluorescence-sourced and calibrated Schlieren measurements, respectively. 
Both investigations identified inclined disturbances extending above the boundary-layer edge, which are quite distinct from the usual rope-like structures associated with nonlinear second-mode instabilities.
Greenwood and Schneider~\cite{greenwood2019measurements} examined the role of entropy-layer instabilities on a~$35\deg$ cone-ogive cylinder at Mach~$6$ and reported that these disturbances grow slowly in the entropy-layer, and, after penetrating the boundary-layer, decrease in magnitude first before showing rapid growth, leading to turbulence onset.
Borovoy et al.~\cite{borovoy2021laminar} investigated the role of various leading edge shapes on transition reversal on blunted plates at Mach~$5$, and reported that, from a practical standpoint, these shapes do not affect transition for nose Reynolds numbers below~$2\times10^{4}$.
Further, at larger nose radii, a reduction of laminar region was reported for a cylindrical leading-edge compared to an elliptical nose.
Similar observations were highlighted in blunt ogive-cylinder experiments by Hill et al.~\cite{hill2021experimental}, where  transition effects were associated with a modification of the entropy-layer structure by the ogive-radius and the nose bluntness.

Linear stability theory (LST) and parabolized stability equations (PSE) have been extensively used in the literature to study flow (de)stabilization due to nose bluntness.
Boundary-layer modal stability investigations on the nose bluntness effects~\cite{jewell2016boundary,lei2012linear,malik1990effect,marineau2014mach} attributed the observed transition trends in low-bluntness regime to a reduction of the boundary-layer edge Reynolds number, and no satisfactory explanation for the observations in medium and large bluntness regimes were apparent.
Zhong and Ma~\cite{zhong2006boundary} examined the receptivity of Mach~$8$ flow over a blunt cone to freestream fast acoustic waves; no evidence was noted of the first-or the second-mode instabilities in the early region of cone surface, despite the local instability of Mack modes. 
Batista et al.~\cite{batista2020mechanism} attributed the suppression of the second-mode to the weakening of the base flow density-gradient, resulting in the disruption of acoustic resonance necessary to sustain its growth.
Hypersonic leading-edge acoustic receptivity studies by Cerminara and Sandham~\cite{cerminara2017acoustic} on blunt wedges predicted that wall response for fast waves results in strong resonant amplification of mode~$F$~\cite{fedorov2011high} and the corresponding response to slow waves exhibits an initial decay and an overall lower amplitude.
He and Zhong~\cite{he2020hypersonic} studied the receptivity effects of freestream slow, fast acoustic, and entropic perturbations on a blunt cone at Mach~$9$ and reported that axisymmetric fast acoustic disturbances generate the largest initial second-mode amplitudes, followed by the axisymmetric hotspot and then the axisymmetric slow acoustic pulse.
Modal amplification of entropy-layer instabilities indicated that the growth-rate of these disturbances were too low to initiate transition~\cite{dietz1999hein, fedorov2004evolution}.
The receptivity studies in the large bluntness regime at Mach~$6$ by Kara et al.~\cite{kara2011effects} further confirmed these observations, where the disturbances initiated upstream of the nose initially decayed by two to three orders of magnitude before starting to grow again due to the second-mode instability.
Despite the significant progress, linear modal instability theory does not provide an adequate understanding of entropy-layer effects.    

Non-modal disturbance growth has recently been proposed as a potential mechanism for the experimentally observed onset of transition on blunt bodies~\cite{reshotko2000blunt,paredes2017blunt,paredes2019nose}.
Significant non-modal amplification of stationary disturbances initiated within the nose-tip vicinity is qualitatively consistent with the observation that transition is highly sensitive to wall roughness in the large-bluntness regime and originates near the nose tip~\cite{stetson1983nosetip}.
Cook et al.~\cite{cook2018understanding} examined the external stability of blunted cones with increasing nose radii and identified the unstable modes appearing in the entropy-layer that dominate  second-mode boundary-layer instabilities. 
Amplification of planar and oblique traveling non-modal disturbances peaking in the entropy-layer has also been reported for blunt cones by Paredes et al.~\cite{paredes2019nonmodal}.
These traveling disturbances can manifest as the experimentally observed inclined structures above the boundary-layer.   
Potential mechanisms for the receptivity of these non-modal waves was provided by Russo et al.~\cite{russo2021particle} by considering particle impingement and Goparaju et al.~\cite{goparaju2021effects} by the interaction of stochastic disturbances with the leading-edge.
Further, Hartman et al.~\cite{hartman2021nonlinear} examined the response of wavepackets initiated in the entropy-layer and reported large growth of \textit{linear} oblique disturbances closer to the leading edge, which were not captured by the conventional linear/parabolic stability theories.  

To clarify the mechanism of turbulence onset through  non-modal amplification of entropy-layer disturbances, Paredes et al.~\cite{paredes2020mechanism} examined the evolution of such oblique disturbances on the frustum of a \textit{medium} bluntness cone. 
Their analysis indicated that such disturbances can interact nonlinearly above the boundary-layer edge, and induce stationary streaks that penetrate the boundary-layer, leading to turbulence onset by streak breakdown.
Furthermore, Hartman et al.~\cite{hartman2021nonlinear} noted the importance of including nonlinear effects to capture relevant transition phenomena observed in high-speed tunnel experiments, and considered the evolution of high-amplitude entropic-instabilities through \textit{controlled} oblique breakdown mechanisms~\cite{fasel1993direct,mayer2011direct,chang1994oblique}.
They identified the progression of transition through the dynamics of positive and negative temperature fluctuations, with the latter penetrating and contaminating the boundary-layer, leading to turbulence onset, while the former remaining largely unchanged. 
This route to transition was also found to be relevant for varying nose bluntness~\cite{meersman2021direct}.

In this work, the evolution of disturbances on blunted \textit{flat-plates} resulting in turbulence onset is studied using high-fidelity numerical techniques.
The nose radius considered is in the \textit{medium} bluntness regime, where, as noted above, although the system supports entropy- and boundary-layer instabilities, modal growth-rates are too-low to initiate transition.
The problem formulation and the pertinent numerical details are outlined in Section~\ref{sec:methodology}.
In Section~\ref{sec:results}, linear stability investigations (\ref{sec:2DLST}) are performed on the laminar base flow to delineate the nature of modal instabilities.
Receptivity of entropy-layer disturbances (\ref{sec:2DDNS}) are next examined with two-dimensional wavetrains actuated at various wall-normal locations and different frequencies.  
The evolution of three-dimensional disturbances (\ref{sec:3DWO}) is also investigated with wavepacket analysis.
A combination of modal decomposition techniques extracts the details of transition to turbulence induced by oblique breakdown of entropic-instabilities (\ref{sec:3DDNS}).
Distinctions in the oblique breakdown mechanism of entropy-layer disturbances in blunted flat-plates versus cones are also highlighted.
Finally, the findings are summarized in Section~\ref{sec:summary}.
Note that, the term \textit{entropic-disturbances} used in this work refers to modal/non-modal perturbations amplifying in the entropy-layer, and not pure entropic fluctuations.

\section{Physical problem and computational techniques}
\label{sec:methodology}
In this section, the problem geometry and simulation set-up are described along with an outline of the operator and data-driven techniques employed for the flow analysis. 
\subsection{Flow parameters}
The geometry and flow conditions (Mach~4) are based on the blunt-nosed flat plate experiments of Lysenko~\cite{lysenko1990influence} and are summarized in table~\ref{tab:param}.
The leading-edge of the geometry is a cylinder of radius~$R_n^{*}=0.5~mm$, and is attached to a flat-plate of length~$l^{*}=400~mm$.
The unit Reynolds number~$(Re_{\infty}^{*})$ of the flow is~$25.3\times 10^{6} /m$, resulting in a nose Reynolds number~$Re_{R_n}=(Re_{\infty}^{*}\times R_n^{*})= 12{,}650$. 
The freestream stagnation temperature is~$T_{0}^{*}~=~290~K$, the wall is isothermal~($T_w^{*}=255~K$) at adiabatic temperature.
\begin{table}
    \centering
    \begin{tabular}{c c c c c c}
     \hline
     \hline
         Parameter & Value &  Parameter & Value & Parameter & Value\\ 
         \hline
          $M_{\infty}$ & $4$ &
          $T_{0}^{*}$ & $290~K$ &
          $l^{*}$ & $400~mm$ \\
          $Re_{\infty}^{*}$ & $25.3 \times 10^{6}~/m$ &
          $R_{n}^{*}$ & $0.5~mm$ &
          $T_w^{*}/T_{\infty}^{*}$ & $3.7$\\
          \hline
          \hline
    \end{tabular}
    \caption{Flow parameters examined in the current numerical investigations, based on the experiments of Lysenko~\cite{lysenko1990influence}. }
    \label{tab:param}
\end{table}

\begin{figure}
    \centering
    \includegraphics[width=0.48\textwidth]{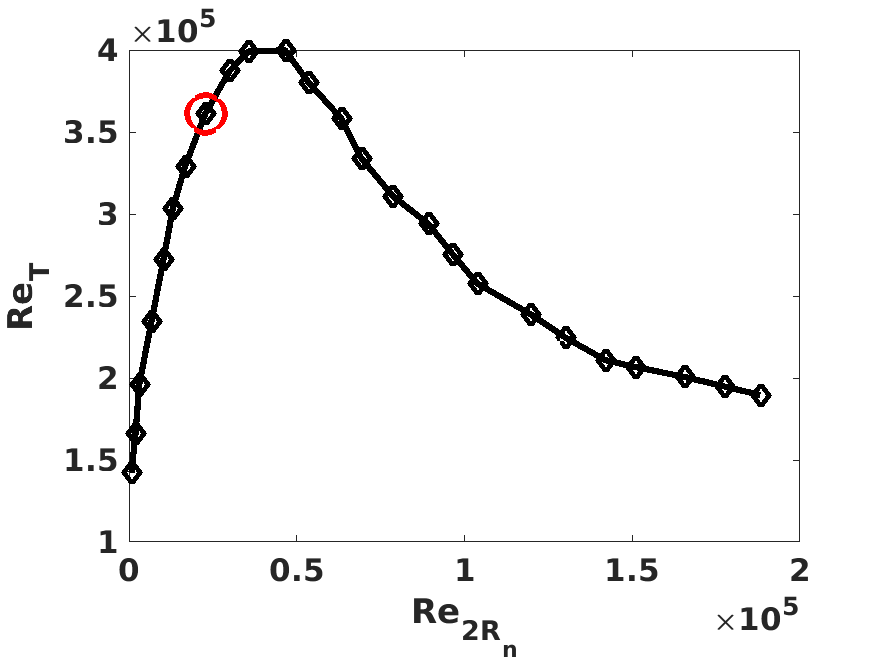}
    \caption{Transition Reynolds number~$(Re_T)$ variation with increasing nose Reynolds number~$(Re_{2R_n})$ depicting transition reversal.
    The nose radius considered in the current work is highlighted.
    Data adapted from Lysenko~\cite{lysenko1990influence}.}
    \label{fig:lysenko}
\end{figure}

The variation of transition Reynolds number~$Re_{T}=Re_{\infty}^{*}\times x_{T}^{*}$ ($x_T^{*}$ is the transition onset location) with the nose Reynolds number, adapted from Lysenko's experiments, is plotted in fig.~\ref{fig:lysenko}.
A downstream movement of transition location with increasing nose radius is observed for $R_n^{*}<1~mm$, followed by its reversal, experimentally capturing transition reversal.
The \textit{medium} bluntness regime of interest is highlighted with red circle.

\subsection{Governing equations}
The flow governing equations are the three-dimensional Navier–Stokes equations.
In conservative form and using Cartesian coordinates, they may be written as:
\begin{equation}
    \frac{\partial U^{*}}{\partial t^{*}}+ \frac{\partial F^{*}_{j}}{\partial x^{*}_{j}} + \frac{\partial F^{*}_{vj}}{\partial x^{*}_{j}} =0, \;(j=1,2,3)
\end{equation}
The tensor notation $(x_1^{*}, x_2^{*}, x_3^{*})$ represents the Cartesian coordinates $(x^{*}, y^{*}, z^{*})$. 
The vector $\textbf{U}^{*}$ contains five conservative flow variables:
\begin{equation}
    \textbf{U}^{*}=[\rho^{*}\quad \rho^{*}u_1^{*}\quad \rho^{*}u_2^{*}\quad \rho^{*}u_3^{*}\quad  e^{*} ]^{T}
\end{equation}
where $t$ is time, $\rho$~is density, $u_{i}$ are the velocity components, and $e$ is internal energy.
Fluxes $F_{j}^{*}$ and $F_{vj}^{*}$ are defined as
\begin{equation}
    F_{j}^{*}=
\left \{
  \begin{tabular}{c}
  $\rho^{*}u_{j}^{*}$\\
  $\rho^{*}u_{1}^{*}u_{j}^{*}+p^{*}\delta_{1j}$\\
  $\rho^{*}u_{2}^{*}u_{j}^{*}+p^{*}\delta_{2j}$\\
   $\rho^{*}u_{3}^{*}u_{j}^{*}+p^{*}\delta_{3j}$\\
  $(e^{*}+p^{*})u_{j}^{*}$
  \end{tabular}
\right \}
\end{equation}
    \begin{equation}
      F_{vj}^{*}=
\left \{
  \begin{tabular}{c}
  $0$\\
  $-\tau_{1j}^{*}$\\
  $-\tau_{2j}^{*}$\\
   $-\tau_{3j}^{*}$\\
  $-\tau_{jk}^{*}u_{k}^{*}-\textbf{q}_{j}^{*}$
  \end{tabular}
\right \} ,\; (k=1,2,3)
\end{equation}
The equations of state and transport are:
\begin{equation}
    p^{*} = \rho^{*}R^{*}T^{*}
\end{equation}
\begin{equation}
\label{eqn:energy}
e^{*}=\rho^{*}(c_v^{*}T^{*}+0.5u_k^{*}u_k^{*})    
\end{equation}
\begin{equation}
    \tau_{ij}^{*}=\mu^{*}(\frac{\partial u^{*}_{i}}{\partial x^{*}_{j}}+\frac{\partial u^{*}_{j}}{\partial x^{*}_{i}})-\lambda^{*}\frac{\partial u^{*}_{k}}{\partial x^{*}_{k}}\delta_{ij}, \; \lambda^{*}=\frac{2}{3}\mu^{*} 
\end{equation}
\begin{equation}
    \textbf{q}_{j}^{*}=-\kappa^{*}\frac{\partial T^{*}}{\partial x_{j}^{*}}
\end{equation}
where $p$ is pressure, $c_v$ is a constant specific heat with a given $\gamma(=1.4)$.
Thermal conductivity $\kappa$ can be determined with a constant Prandtl number $(Pr=0.72)$ along with the viscosity coefficient~$\mu$ defined by Sutherland’s law.

\subsection{Numerical methodology and boundary conditions}
A high-order shock-capturing method is employed to accurately resolve the features of the leading bow shock and the fine-scale structures generated during the laminar-turbulent transition. 
The Navier–Stokes equations in Cartesian coordinates $(x^{*}, y^{*}, z^{*}, t^{*})$ are transformed into body-fitted curvilinear computational domain coordinates $(\xi, \eta, \zeta, \tau)$.
The governing (dimensional) equations are non-dimensionalized using the scales $l_{nd}^{*}=200~mm$ for lengths, ~$u_{\infty}^{*}$ for velocities, $\rho_{\infty}^{*}$ for densities, ~$\rho_{\infty}^{*}u_{\infty}^{*2}$ for pressure and ~$l_{nd}^{*}/u_{\infty}^{*}$ for time. 
The frequency is denoted by the Strouhal number~$(St=f^{*}l_{nd}^{*}/u_{\infty}^{*})$ and the results hereafter are presented in nondimensional form (without asterisk).
Note that $(x,y,z)$ denote streamwise, wall-normal and spanwise directions, and $(u,v,w)$ are the corresponding velocity components.

The transition investigation presented here follows the strategy employed in the earlier studies by Fasel and co-workers~(see,~\cite{hartman2021nonlinear,hader2019direct}).
The first step is to simulate the two-dimensional laminar base flow over a blunted flat plate, including the leading edge, with a low-order scheme on a relatively coarse grid.
To this end, the inviscid spatial flux derivatives in the streamwise and wall-normal directions are obtained using the Roe~\cite{roe1981approximate} scheme with third-order reconstruction.
The van Leer harmonic limiter~\cite{van1979towards} is employed to damp grid-scale oscillations.
A second-order central finite difference scheme is used for the spatial derivatives of the viscous flux in the streamwise and wall-normal directions. 
Time marching is performed with an explicit, nonlinearly stable third-order Runge–Kutta scheme~\cite{shu1988efficient}.
For the boundary conditions, the surface of the wall is assumed to be no-slip, isothermal at the recovery temperature noted in table~\ref{tab:param}.
Zero gradient conditions are applied at the nose symmetry plane and the exit.
Freestream conditions $u=1$, $v=0$, $\rho=1$, $p=1/(\gamma M_{\infty}^{2})$ are specified at the top boundary.

After the base flow is converged, in the second step, the nose segment is removed and the solution is re-interpolated on a finer grid with Dirichlet boundary conditions at the inlet, with the top boundary conditions remaining the same.
At the exit plane, the grid is stretched downstream of the domain of interest to damp the acoustic reflections.
The finer grid has~$3{,}000\times400$ points in streamwise and wall-normal directions, uniformly spaced and exponentially stretched, respectively. 
The near-wall spacing is~$\Delta y=5\times 10^{-5}$, corresponding to~$y^{+}\sim 0.8$ wall grid spacing.
The streamwise spacing ensures a minimum of~$50$ points per streamwise wavelength of the fundamental forcing wave considered, comparable to earlier transition studies~\cite{paredes2020mechanism}.
Higher-order numerical techniques are employed in these resolved simulations as follows.
A seventh-order weighted essentially non-oscillatory (WENO) reconstruction of characteristic variables~\cite{balsara2000monotonicity} is used in conjunction with the Roe Riemann solver.
The scheme degenerates to a fifth-order WENO at the third point from the boundary and a monotonic upwind scheme for conservation laws (MUSCL) reconstruction at the second and first points.
Sufficient overlap is included at the processor block boundaries to preserve the high-order accuracy.
To increase the stability near shocks, a Ducros~\cite{ducros1999large}-like sensor is employed and the inviscid scheme is locally switched to a low-order Riemann-solver. 
The viscous fluxes are calculated using a sixth-order, compact-central-difference, spectral-like scheme~\cite{lele1992compact,visbal2002use}.
Time is discretized using a variant of the implicit, approximately factored, second-order Beam and Warming method~\cite{beam1978implicit} in a diagonalized form~\cite{pulliam1981diagonal}. 
A nondimensional time step of~$\Delta t=5\times10^{-5}$ is used, and three Newton-like sub-iterations are performed in the implicit scheme at each time step to recover the accuracy due to errors introduced by linearization, factorization and explicit updating of the boundary conditions.

For the linear perturbation evolution studies, the inflow and the outflow of the domain are located at $x=0.3$ and $x=1.95$, respectively.
For the nonlinear breakdown studies, this extent is chosen based on the wavelengths of the amplifying disturbances considered.
The $3D$ wavepacket analyses are performed with a domain extruded (using $201$ points) in the spanwise direction in the range~$z \in [-0.2,0.2]$ to ensure that the disturbances do not reflect at the (lateral) boundary.
Spanwise periodic conditions are imposed in both cases.
The disturbance forcing functions employed for the wavepacket and breakdown studies are now discussed.

To examine the receptivity and instability evolution, perturbations are imposed at various wall-normal locations as shown schematically in fig.~\ref{fig:schematic}. 
Evolution of $2D$ and $3D$ disturbances are investigated through pulsed and continuous forcing.
For the wavepacket analysis, disturbances are seeded in the internal energy~$(e)$~(see eq.~\ref{eqn:energy}), through a monopole spatial forcing of the form eq.~\ref{eqn:forcing1} at~$z=0$~\cite{novikov2016direct}.
\begin{equation}
    \begin{aligned}
    e' &= 0  & t < t_{1} \\
       & =A \sin(\pi \frac{x-x_1}{x_2-x_1}) \sin(\pi \frac{y-y_1}{y_2-y_1}) \sin(-\omega t) & t_{1}\le t \le t_{2} \\
       & =0 & t \ge t_{2}
    \end{aligned}
    \label{eqn:forcing1}
\end{equation}
where $A$ is the amplitude, $(x_i,y_i)$ are the extents of streamwise, wall-normal forcing location.
Also,~$(\new{t_{1}=0},t_{2})$ are the start and the end of the temporal pulse, with~$\omega=2\pi St=2 \pi /(t_{2}-t_{1})$.
For wall forcing, the wall-normal component~$(y)$ in eq.~\ref{eqn:forcing1} is replaced by blowing-suction (in~$z$) and for the other cases, the disturbances are imposed at entropy- or boundary-layer edges.
For continuous forcing cases, the disturbance form is given by:
\begin{equation}
    \begin{aligned}
    e' =\Sigma_{n=1}^{N} A_{n} \sin(\pi \frac{x-x_1}{x_2-x_1}) \sin(\pi \frac{y-y_1}{y_2-y_1}) \sin(-\omega_{n} t+\beta_{n} z) 
        \end{aligned}
    \label{eqn:forcing2}
\end{equation}
where~$\beta_{n}$ is the $n^{th}$~spanwise wavenumber.
While~$\beta=0$ for $2D$-simulations, two oppositely oriented spanwise waves can be specified by the summation of two waves~$(N=2)$ with~$\beta_1=\beta$, $\beta_2=-\beta$. 
The monopole width is~$0.025$ in the continuous and wavepacket forcing investigations.
The linear $2D$ and wavepacket analyses were performed with an amplitude~$A=1\times 10^{-4}$, and the $3D$ continuous forcing was triggered with an amplitude~$9\times 10^{-2}$ to study the nonlinear interactions.  
\new{The higher amplitudes employed for the nonlinear case ensures breakdown in the domain chosen based on the experiments of Lysenko~\cite{lysenko1990influence}, and are commensurate with the forcing amplitudes used in earlier numerical experiments of Leinemann et al.~\cite{leinemann2021direct}, Franko and Lele~\cite{franko2013breakdown}.}

\new{The energy is chosen as the actuation variable based on earlier experimental~\cite{chou2010characterization} and numerical~\cite{hartman2021nonlinear} investigations.
Other variables could also be used for modeling freestream disturbances, such as vortical gusts (through velocity components) or freestream acoustic waves, which could vary the receptivity mechanism~\cite{cerminara2020transition}.
These effects are not considered in this paper.
}
The cases considered in this study are summarized in table~\ref{tab:sim_cond}.
\begin{table}
    \centering
    \begin{tabular}{c c c}
     \hline
     \hline
         Dimensionality & Forcing type & Forcing location\\ 
         \hline
         &  & wall \\
         $2D$ linear  & continuous & boundary-layer edge \\
              &  & entropy-layer edge \\\hline
          &  & wall \\
        $3D$ linear  & wavepacket & boundary-layer edge \\
              &  & entropy-layer edge \\\hline
         $3D$ non-linear & continuous & entropy-layer edge \\
          \hline
          \hline
    \end{tabular}
    \caption{Summary of numerical simulations performed in the current study.}
    \label{tab:sim_cond}
\end{table}

\subsection{Local stability analysis}
One-dimensional linear stability calculations are conducted to characterize the unstable modes of the boundary- and  entropy-layers.
These have proven useful in informing the linear stages of laminar-turbulent breakdown in hypersonic configurations with sharp leading edges~\cite{unnikrishnan2020linear, hader2019direct, franko2013breakdown}.
As noted earlier, LST has also been employed to account for the destabilization of modal instabilities in the entropy-layer~\cite{fedorov2004evolution, dietz1999hein}. 

In LST, the flow variables~$q=[u\ v\ w\ p\ \rho]^{T}$ are decomposed into steady (laminar) reference state~$\bar{q}$ and an unsteady small perturbation field~$q'$.
The reference state is assumed to be locally parallel in the streamwise direction and the perturbation is a function of~$y$, resulting in the ansatz:
\begin{equation}
\label{eqn:lst}
    q'(x,y,z,t)=\hat{q}(y)\exp[i(\alpha x+\beta z-\omega t)]+c.c
\end{equation}
where~$\hat{q}$ is a one-dimensional amplitude function,~$i$ is square root of unity, and~$c.c$ is the complex conjugate.
A temporal formulation is employed to solve the stability equations; $\omega$ is thus a complex frequency,~$(\alpha, \beta)$ are the real streamwise and spanwise wavenumbers.
While the real part of~$\omega$ is the frequency~$(\omega_r)$, its imaginary part~$(\omega_i)$ indicates growth-rate, and denotes amplification/damping if it is positive/negative.

To perform the stability calculations, the reference laminar DNS base state is re-interpolated onto a Chebyshev grid by means of a cubic spline interpolation in the wall-normal direction.
Dirichlet boundary conditions are used at the wall and the freestream, with the latter extending above the entropy-layer edge.
Substituting the ansatz~(eq. \ref{eqn:lst}) into the linearized Navier-Stokes equations yields a generalized eigenvalue problem~(see~\cite{malik1990numerical}), which is solved using QR decomposition in MATLAB.
Note that while the length in LST calculations is non-dimensionalized by the Blasius scale, for consistency, the later results are reported using DNS nondimensional scales discussed previously.

\subsection{Data-driven flow decomposition techniques}
Coherent structures of complex fluid systems facilitate the identification of a low-dimensional representation that captures the prominent dynamics of interest.
The choice of technique and the flow variable(s) subjected to the analysis depends on the purpose.
Three data-driven flow decomposition techniques are ideally suited for the goals identified in Section~\ref{sec:intro}.

Dynamic mode decomposition (DMD), which is based on Koopman analysis of nonlinear dynamical systems~\cite{mezic2013analysis}, is employed to unveil the coherent structures at a \textit{single-frequency}.
It extracts a reduced-order description of a linear transformation that maps any snapshot from the data sequence onto the subsequent snapshots.
This technique has been extensively applied to study dynamics of the transitional flow field in the literature~\cite{sayadi2014reduced,bucci2021influence}.
For spatio-temporally varying snapshots~$L$, DMD decomposition can be defined as
\begin{equation}
L(x,y,z,t)=\Sigma_{n=1}^{N}a_n\phi_n(x,y,z)\psi_n(t)
    \label{eqn:dmd}
\end{equation}
where~$\phi_n$ is the~$n^{th}$ spatial mode whose amplitude is~$a_n$.
The temporal variation in~$\psi_n(t)$ consists of a single complex frequency,~$\lambda =\lambda_r +i\lambda_i$, where~$(\lambda_r, \lambda_i)$ are growth-rate and frequency, respectively.
While DMD is equivalent to a global stability analysis in a linearized flow simulation, for nonlinear data sequences it describes the most dominant dynamic behaviour~\cite{sayadi2014reduced}.
In this work,~$3D$ streamwise velocity snapshots are processed using DMD algorithm of Schmid~\cite{schmid2010dynamic}.
\new{The disturbances are extracted by subtracting the average of the data sets (i.e. mean flow) rather than the laminar base flow.
The former approach is closely related to Fourier analysis and good convergence is achieved with fewer snapshots through the construction of a best-fit linear operator~\cite{towne2018spectral}.
On the other hand, subtracting laminar base flow allows DMD to adjust frequencies to capture the dynamics not resolved by the flow data sets~\cite{chen2012variants}.
}

Linear instabilities (modal/non-modal) provide mechanisms for disturbance growth, and upon saturation, nonlinear interactions redistribute the energy among disturbance frequencies and thus have no net effect on the instantaneous energy growth-rate~\cite{drazin2004hydrodynamic}.
Wave resonances play a vital role in this energy transfer: a detailed review of various such mechanisms is provided by Wu~\cite{wu2019nonlinear}. 
While DMD/Fourier transformation are applicable to linear and nonlinear flow data, they do not explicitly account for the nonlinear interactions.
Higher-order statistical tools, such as auto- and cross-bispectrum, have been employed in the literature to quantify some of these interactions in experimental~\cite{hill2021experimental} and numerical~\cite{unnikrishnan2020linear} studies.
Auto-bispectrum reveals coupled nonlinear interactions in the same variable, while cross-bispectrum identifies interactions of multiple variables.
Although the bispectrum highlights quadratic phase coupling, it is limited to one-dimensional signals and cannot establish the causality among the interacting waves.  
To overcome these limitations, the current work employs bispectral mode decomposition~(BMD, \cite{schmidt2020bispectral}) of multidimensional data to educe flow structures associated with triadic interactions.
Here,\todo{This is a bit of a complicated sentence.. may want to break it up} the integral measure of a third-order statistic, bispectral density, $S_{qrs}=\lim_{t \rightarrow \infty}\frac{1}{t}E[\hat{q}(St_1)\hat{r}(St_2)\hat{s}(St_1+St_2)^{*}]$ is maximized to quantify the (\textit{cause})
interactions of the variable~$q$ at (frequency)~$St_{1}$ with variable~$r$ at~$St_2$, resulting in (\textit{effect}) energy transfer to variable~$s$ at~$(St_1+St_2)$.
Note that~$E$ is the expectation operator with time~$(t)$ extending to infinity~$(\infty)$, the hat~$(\wedge)$ denotes amplitude in spectral-space and the asterisk~$(*)$ is the complex conjugate.
The method is termed auto or cross bispectral modal decomposition (CBMD) if the variables~$(q,r,s)$ are the same or different, respectively, consistent with the definition of bispectrum.
CBMD also extracts an \textit{interaction map}~$\psi$, that quantifies the average local bicorrelation among the interacting frequencies, indicating the regions of triadic activity. 
The cross-bispectrum characterizing the interactions of temperature perturbations~$(q=T,r=T)$ resulting in transfer of energy to streamwise velocity perturbations~$(s=u)$ are examined in the current study, utilizing the algorithm outlined by Schmidt~\cite{schmidt2020bispectral}.

\section{Results}
\label{sec:results}
\subsection{Steady base flow and linear stability analysis}
\label{sec:2DLST}
\begin{figure}
    \centering
    \includegraphics[trim=300 40 320 900,clip,width=0.8\textwidth]{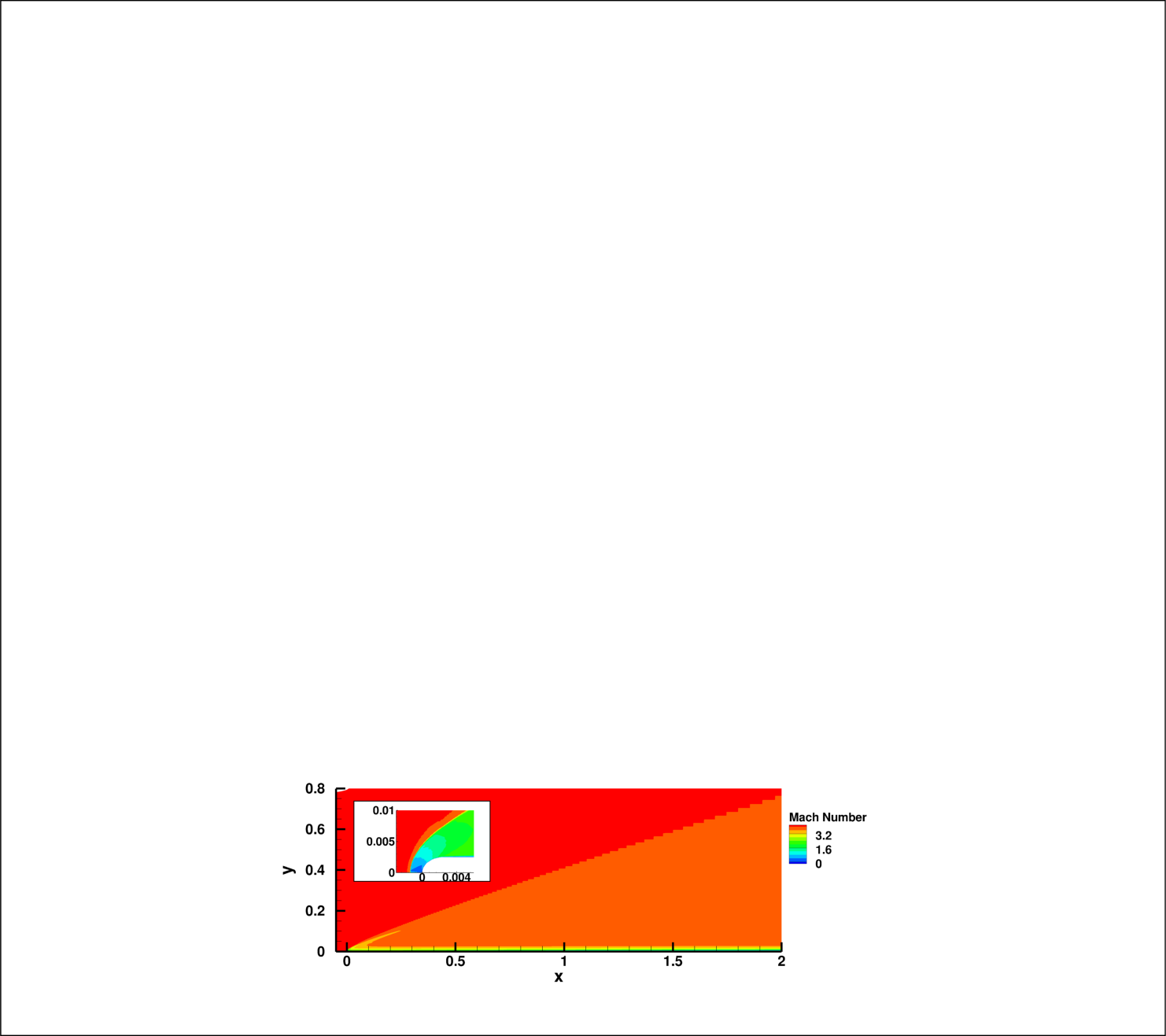}
    \caption{Mach number contours of the steady laminar base flow~($R_n^{*}=0.5~mm$).
    Nose region is highlighted in the inset.}
    \label{fig:Mach}
\end{figure}

The development of steady base flow on the blunted flat plate is first discussed, followed by linear stability investigations.
Mach number contours of the steady base flow over the blunted plate at Mach~$4$ are shown in fig.~\ref{fig:Mach}, with the leading-edge portion highlighted in the inset.
The accuracy of the steady base flow is crucial for transition studies to effectively capture the evolution of small amplitude perturbations. 
The leading-edge bow shock is a vital feature of this flow field and hence empirical relations are employed to ensure the correctness of the shock-standoff distance.
A good match is obtained between the numerical shock stand-off distance~$(\delta = 0.00135)$ and the corresponding empirical correlation of Sinclair and Cui~\cite{sinclair2017theoretical} $(\delta = 0.00150)$.
The initially strong normal shock at the symmetry plane weakens as it curves around the plate and ultimately evolves into a Mach wave.
As noted earlier, the turning of the shock at the leading-edge generates an entropy gradient across it, resulting in a strong entropy-layer around the body.

\begin{figure}
\centering
\begin{subfigmatrix}{2}
\subfigure[]{{\includegraphics[width=0.49\textwidth]{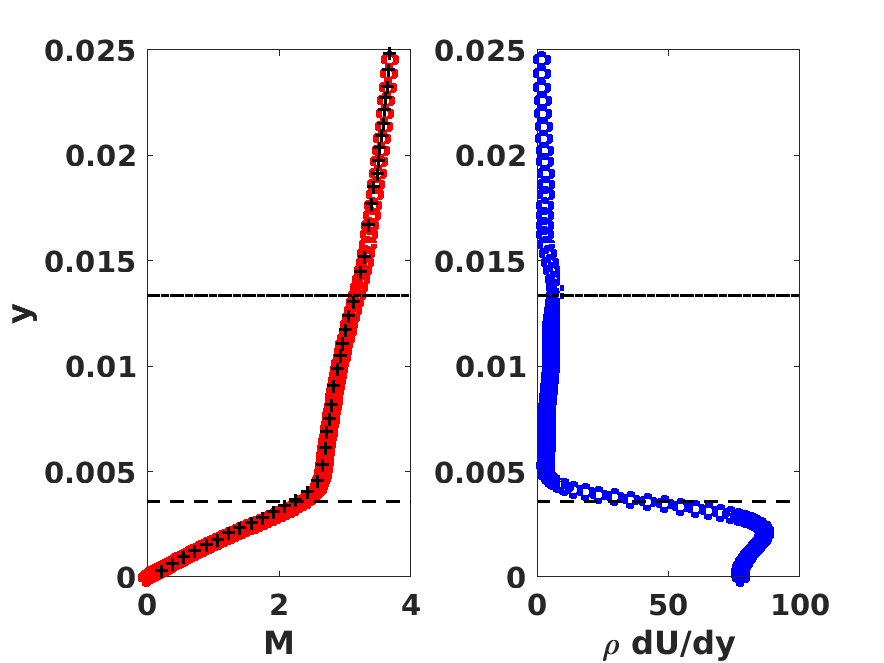}}}
\subfigure[]{{\includegraphics[width=0.49\textwidth]{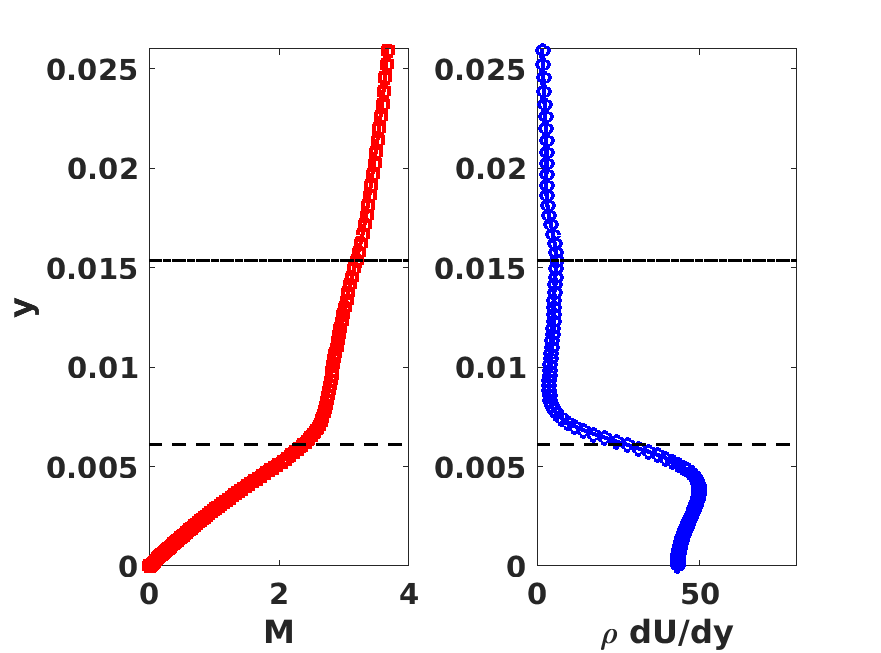}}}
\end{subfigmatrix}
\caption{Mach number~($M$) and generalized inflection profile~($\rho dU/dy$) at: $(a)~x=0.4$ $(b)~x=1.2$.
\new{The~$+$ symbols in $(a)$ denotes Mach profile extracted from Scholten et al.~\cite{scholten2022linear}.}
Boundary- and entropy-layer edges are marked by dashed~$(- -)$ and dash-dot~$(- .)$ lines.}
\label{fig:GIP}
\end{figure}

Boundary- and entropy-layer thicknesses are vital parameters governing the evolution of instability waves.
For a consistent definition, the boundary-layer thickness~($\delta_h$) is established based on the total enthalpy as the first location where its wall-normal gradient is negative.
The entropy-layer edge~($\delta_s$) is defined using entropy $S$ at the wall-normal location where $S-S_{\infty}=0.25\times(S_{w}-S_{\infty})$~\cite{paredes2019nose}, with $S_w$, $S_{\infty}$ representing the corresponding wall and freestream values.
Figure~\ref{fig:GIP}(a,b) shows wall-normal Mach number and generalized inflection profiles~$(\rho dU/dy)$ at the streamwise locations $x=0.4$,~$1.2$.
\new{Good agreement with the Mach profile extracted from Scholten et al.~\cite{scholten2022linear} is found, validating the base flow with VULCAN solver~\cite{litton2003algorithmic}.}
The boundary- and entropy-layer edges are denoted by dashed~$(- -)$ and dash-dot~$(- .)$ lines, respectively, and this convention is consistent with other plots in this paper.
Increases in the boundary-layer~$(y \sim 0.0040$-$0.0061$) and entropy-layer~($y \sim 0.0137$-$0.0154$) thicknesses are observed in the downstream direction.
Trends of entropy-layer swallowing are also apparent, with the entropy-layer swallowing point located outside the current domain of interest.
The generalized inflection point~$d(\rho dU/dy)/dy = 0$ (GIP) occurs at $y\sim 0.0025$,~$0.0039$ in fig.~\ref{fig:GIP}(a,b), indicating the existence of inviscid instabilities in the boundary-layer.
Similarly, the entropy-layer is also susceptible to modal instabilities due to the inflection points at $y\sim 0.0013$,~$0.0015$~\cite{dietz1999hein}.

\begin{figure}
\centering
\begin{subfigmatrix}{2}
\subfigure[]{{\includegraphics[trim=5 0 18 0,clip,width=0.49\textwidth]{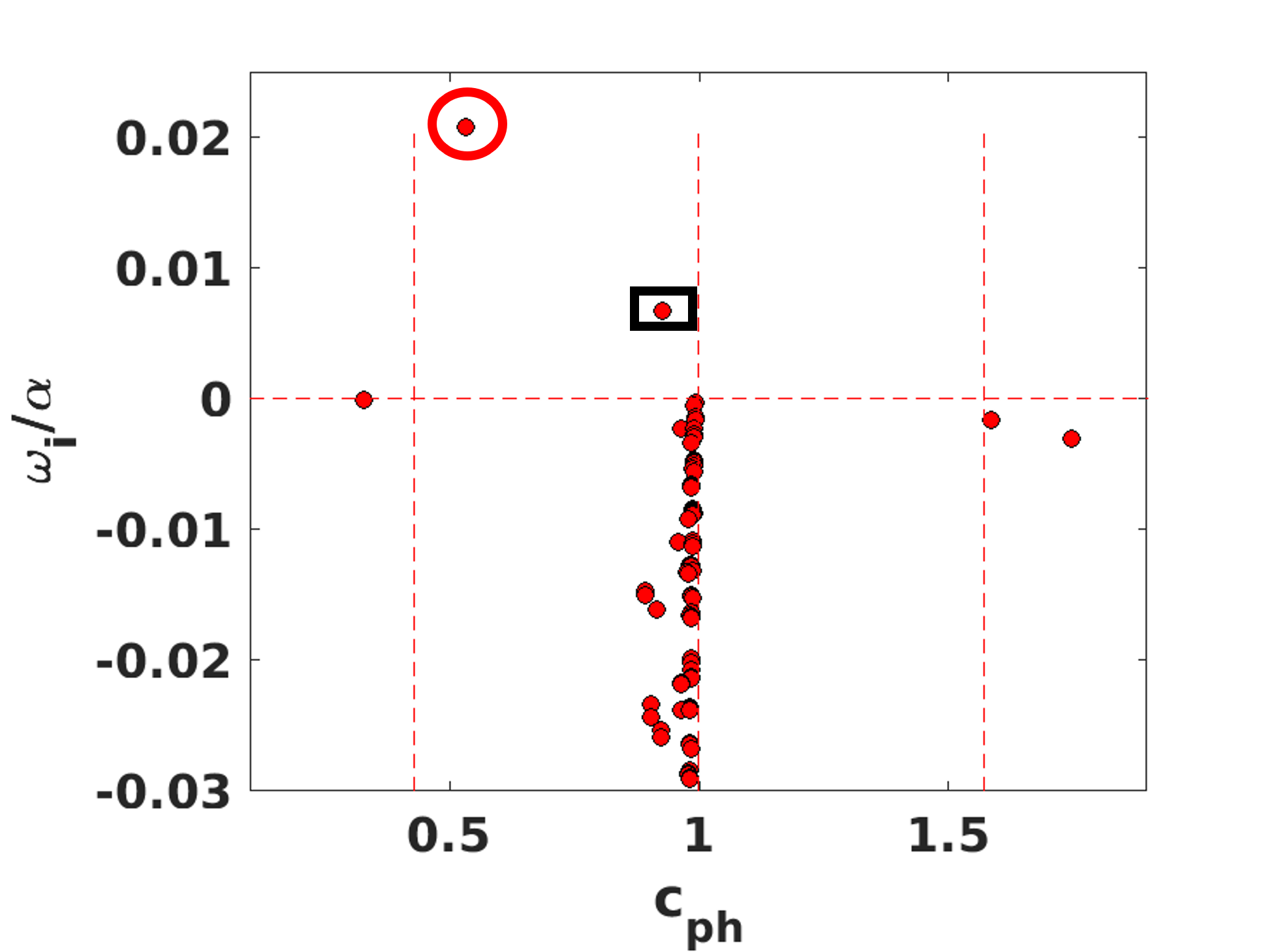}}}
\subfigure[]{{\includegraphics[trim=5 0 18 0,clip,width=0.49\textwidth]{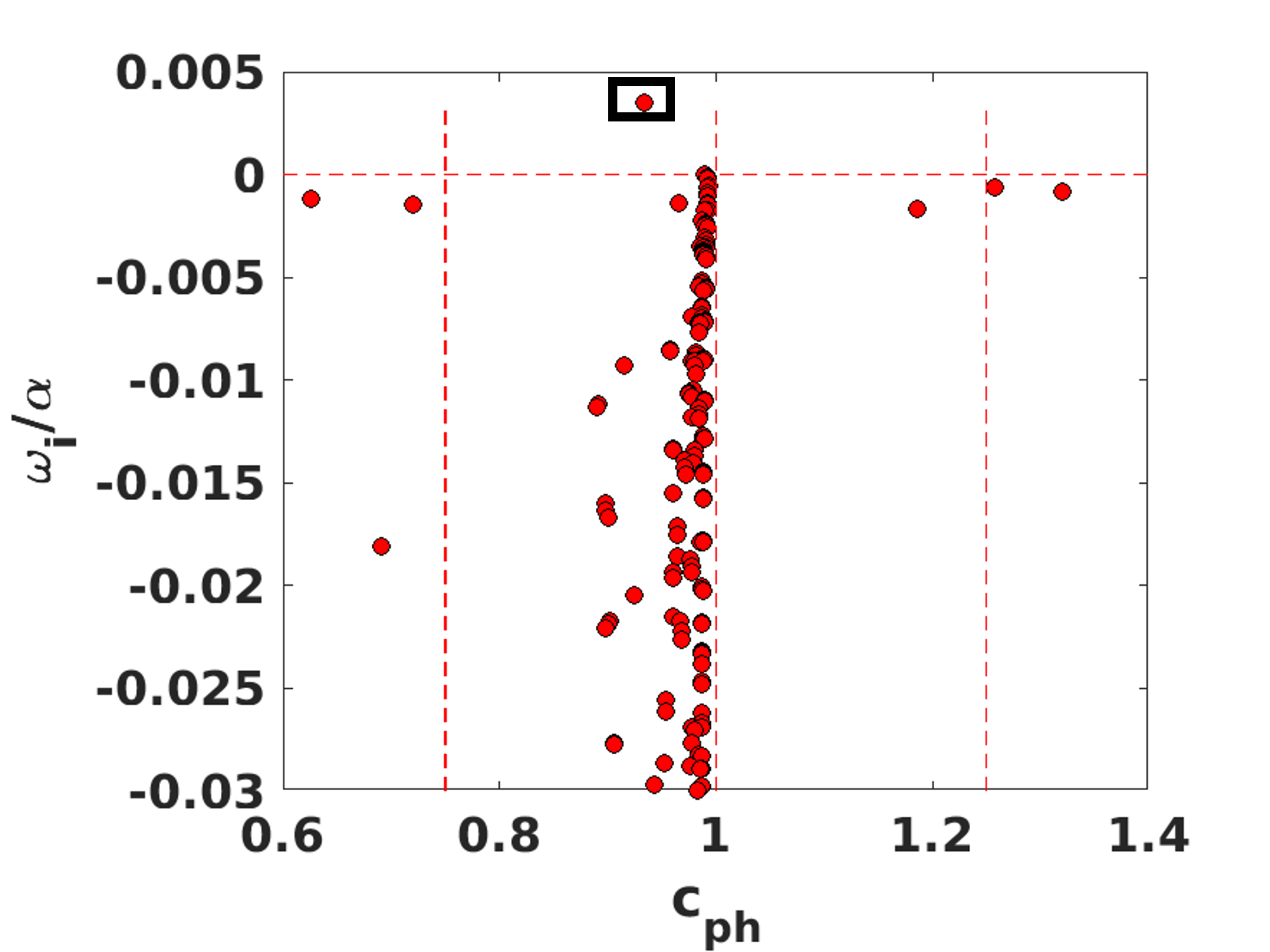}}}
\end{subfigmatrix}
\caption{Eigenspectrum from linear stability theory at~$x=1.2$.
$(a)$~$(\alpha,\beta)=(34.08,70)$ $(b)$~$(\alpha,\beta)=(131.63,0)$.
The (red) circle and (black) rectangle highlights unstable boundary- and entropic-instabilities, respectively.}
\label{fig:LST}
\end{figure}

Modal instabilities in the boundary- and entropy-layer are now examined at a typical streamwise location~$(x=1.2)$. 
Figure~\ref{fig:LST} highlights the complex eigenvalues (phase speed~$(c_{ph}=\omega_r/\alpha)$ and temporal growth-rate~$(\omega_i/\alpha)$) obtained from the LST analysis.
The vertical line~$c_{ph}=1$ denotes the vorticity and entropy spectra, and $c_{ph}=1\pm \frac{\sqrt(1+\beta^2/\alpha^2)}{M_{\infty}}$ are the fast/slow acoustic limits~\cite{forgoston2005initial}.
At~$\alpha=34.08$, while oblique boundary-layer modes were unstable ($\beta=70$ shown), negligible amplification of~$2D$ first-mode waves was found.
In fig.~\ref{fig:LST}(a), this unstable mode is highlighted with a red circle at $(c_{ph},\omega_i/\alpha)=(0.533,0.0208)$ and $St=2.89$.
Another mode, closer to continuous spectra, at $(c_{ph},\omega_i/\alpha)=(0.927,0.0067)$ and $St=5.03$ is also amplified (black rectangle), and is the entropic-instability associated with the entropy-layer GIP in fig.~\ref{fig:GIP}(b).
Figure~\ref{fig:LST}(b) depicts two-dimensional entropic-instability at a higher frequency~$(St=19.55)$, where no unstable ($2D/3D$) boundary-layer instabilities are found (not shown).
\new{The highlighted mode has $(c_{ph},\omega_i/\alpha)=(0.9336,0.00349)$, and the phase speed qualitatively matches with the entropic-instabilities~$(c_{ph}=0.9325)$ reported by Fedorov and Tumin~\cite{fedorov2004evolution} on blunt plates at Mach~$6$.}

\begin{figure}
\centering
\begin{subfigmatrix}{2}
\subfigure[]{{\includegraphics[trim=90 0 110 0,clip,width=0.3\textwidth]{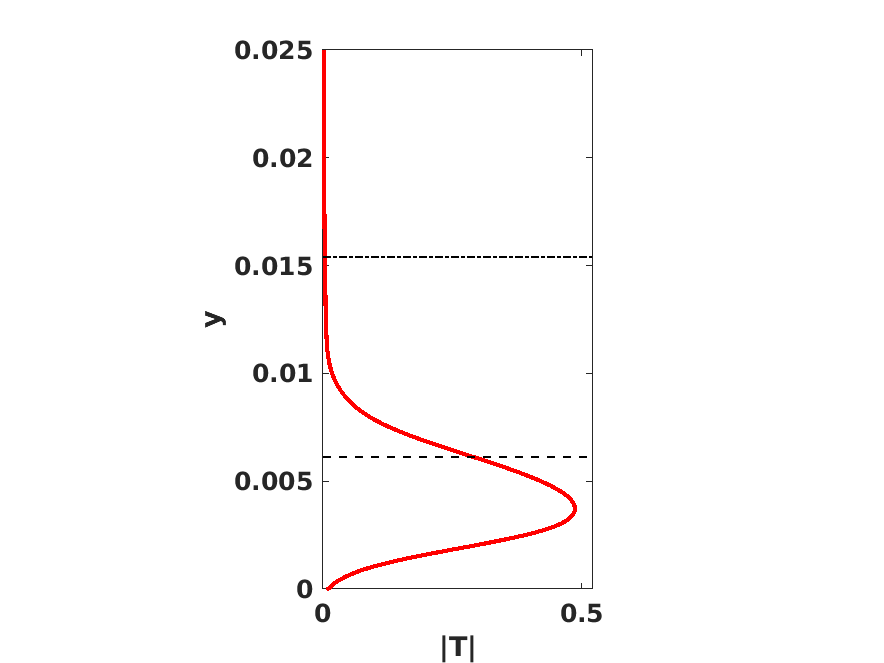}}}
\subfigure[]{{\includegraphics[trim=90 0 110 0,clip,width=0.3\textwidth]{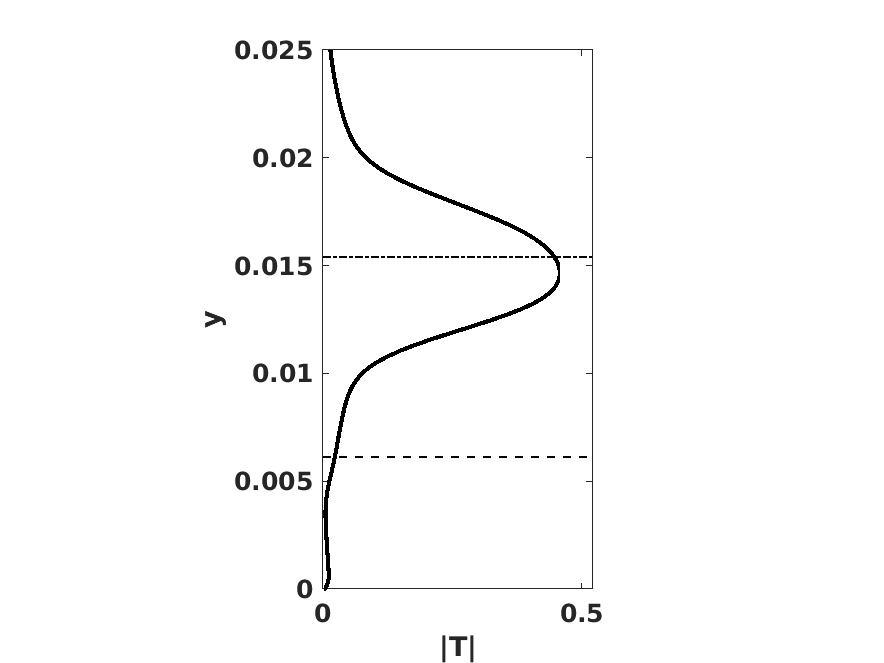}}}
\end{subfigmatrix}
\caption{Temperature eigenstructure of unstable modes.
Amplitude of $(a)$~Boundary-layer instability at~$St=5.03$, $\beta=70$ $(b)$~Entropy-layer instability at~$St=19.55$, $\beta=0$.
}
\label{fig:Mode_LST}
\end{figure}

The corresponding wall-normal structures of the temperature eigenfunctions of the boundary- and entropy-layer instabilities are depicted in fig.~\ref{fig:Mode_LST}.
In fig.~\ref{fig:Mode_LST}(a) the oblique boundary-layer mode highlighted in fig.~\ref{fig:LST}(a) is plotted.
The peak of the eigenfunction is close to the boundary-layer inflection point and closely matches that of a typical first-mode.
The high-frequency entropic-instability at $St~=~19.55$ in fig.~\ref{fig:Mode_LST}(b) peaks in the entropy-layer.
\new{As expected, these features qualitatively agree with the entropic-instability profiles reported by Fedorov and Tumin~\cite{fedorov2004evolution}.}
Thus, the current LST analysis reinforces earlier studies confirming the existence of entropic-instabilities.
Nevertheless, their growth-rates are too low to initiate the transition through a modal amplification mechanism alone. 
Next, we numerically examine the receptivity and the nonparallel effects of these disturbances.

\subsection{Linear evolution of continuous two-dimensional perturbations}
\label{sec:2DDNS}
\begin{figure}
\centering
\begin{subfigmatrix}{3}
\subfigure[]{{\includegraphics[trim=120 0 140 0,clip,width=0.95\textwidth]{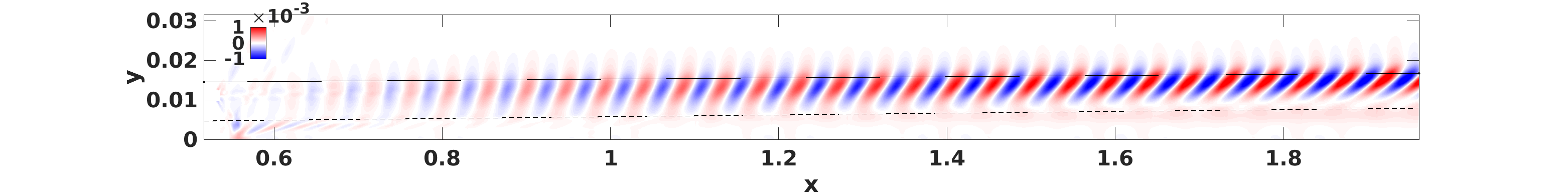}}}
\subfigure[]{{\includegraphics[trim=120 0 140 0,clip,width=0.95\textwidth]{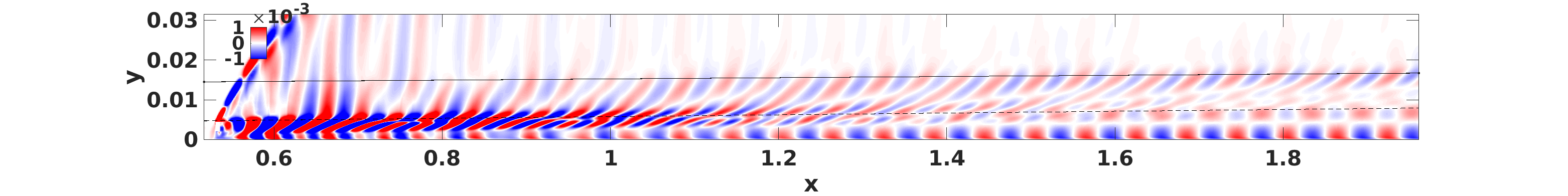}}}
\subfigure[]{{\includegraphics[trim=120 0 140 0,clip,width=0.95\textwidth]{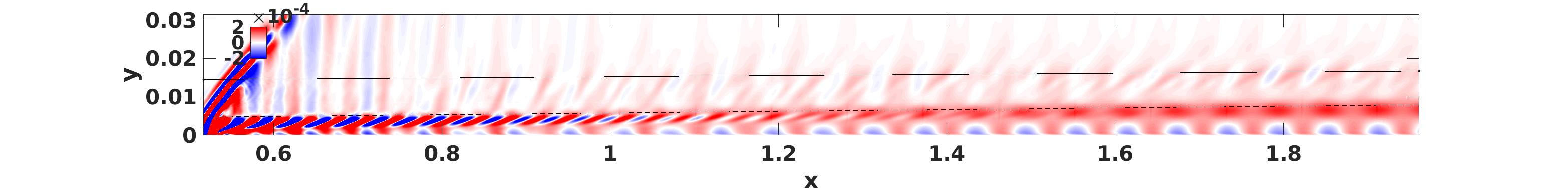}}}
\end{subfigmatrix}
\caption{Contours of instantaneous temperature perturbations illustrating effects of the actuator location.
Forcing at $(a)$~Entropy-layer edge $(b)$~Boundary-layer edge, and $(c)$~Wall.
Entropy- and boundary layer-edges are marked by dotted-dashed~$(.-)$ and dashed~$(--)$ lines, respectively.
}
\label{fig:rhop}
\end{figure}

To examine the evolution of $2D$ disturbances, wavetrains at multiple frequencies are initiated at various wall-normal locations.
The streamwise variation of the instantaneous temperature perturbation field forced at a frequency~$St=20$ is illustrated in fig.~\ref{fig:rhop}.
Signatures of the forced perturbations translating into the entropy-layer, and reflecting from the wall are apparent near the forcing location~$(x\sim 0.5)$.
\new{The radiating disturbances at $x\sim 0.5$ are associated with actuator inputs and do not affect the receptivity process~(see, \cite{unnikrishnan2019interactions}).}
Coherent structures amplify above the boundary-layer (fig.~\ref{fig:rhop}(a)) with the perturbations seeded in the entropy-layer.
This is characterized by an increased amplitude closer to the entropy-layer edge and stretching in wall-normal direction.
Due to the wall-normal flow gradients, the fluid closer to the boundary-layer convects slower, resulting in streamwise tilted structures.
At~$x=1.2$, the disturbance wavelength is~$\lambda_x=0.0455$, resulting in a phase speed~$c_{ph}(=\lambda_x\times St)$ of~$0.9108$, closely matching that of entropic-instabilities.
In fig.~\ref{fig:rhop}(b,c) disturbance evolution signature corresponds to the forcing at boundary-layer edge, and wall, respectively.
While there is a signature of waves propagating in the entropy-layer, no significant amplification and coherent structure can be inferred.
Roller-like structures noticed near the wall are purely translating phenomena and exhibit no growth-rates, consistent with the earlier discussion on linear instability.
Hence, entropic-instabilities are most receptive to perturbations triggered in the entropy-layer.
For blunt cones at Mach~$8$, Husmeier and Fasel~\cite{husmeier2007numerical} also reported that the flow field was most receptive to the disturbances forced in the entropy-layer.

\begin{figure}
    \centering
    \begin{subfigmatrix}{2}
    \subfigure[]{{\includegraphics[trim=0 0 0 0, clip, width=0.6\textwidth]{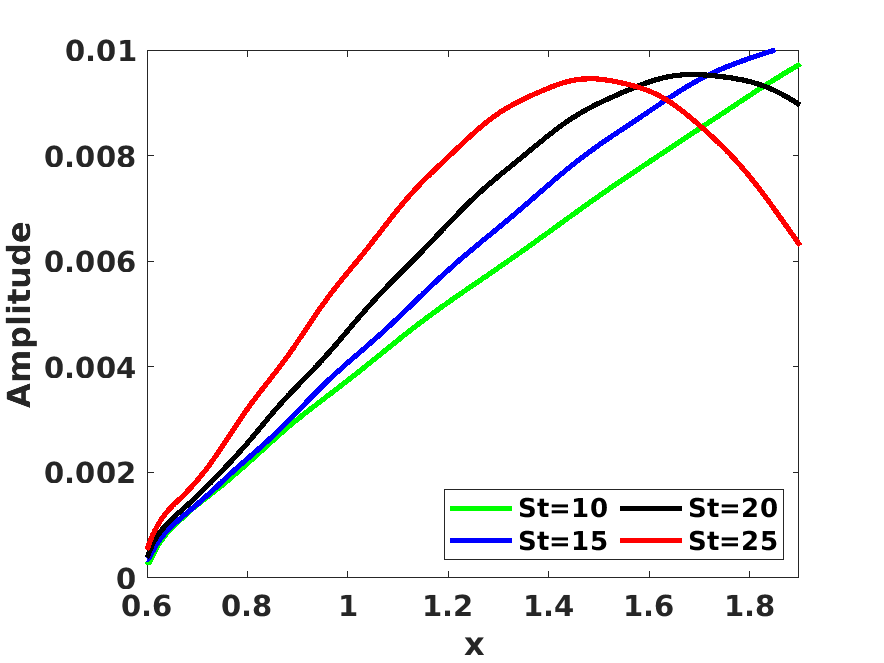}}}
    \subfigure[]{{\includegraphics[trim=0 0 220 0, clip, width=0.3\textwidth]{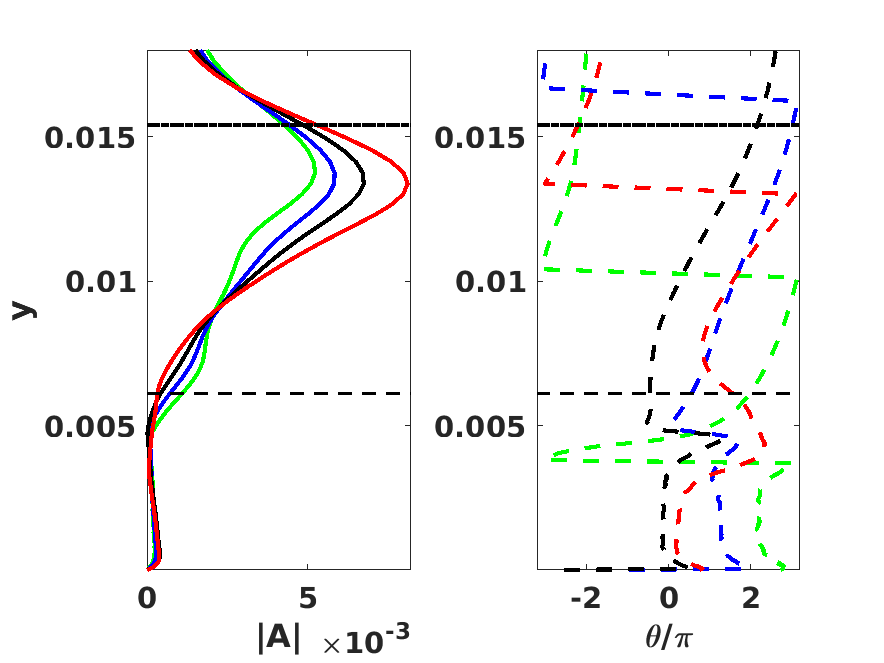}}}
    \end{subfigmatrix}
    \caption{Features of two-dimensional temperature perturbation evolution seeded in the entropy-layer.
    $(a)$ Amplitude for various frequencies at the wall-normal height~$y=0.014$.
    $(b)$ Wall-normal variation of the amplitude~$|A|$ extracted from Fourier transform of temperature fluctuations at the streamwise location~$x=1.2$.}
    \label{fig:2D_Amp}
    \end{figure}

Disturbance evolution features of the wavetrains seeded in the entropy-layer at various frequencies are highlighted in fig.~\ref{fig:2D_Amp}. 
In fig.~\ref{fig:2D_Amp}(a), the streamwise evolution of temperature perturbation amplitudes extracted at a wall-normal height~$y=0.014$ are plotted.
The higher frequencies have larger amplification-rates, followed by an earlier decay.
For~$St=20$ at~$x=1.2$, the local growth-rate~$(-\frac{1}{A}\frac{dA}{ds})$, ($A$ is amplitude) is~$1.25$, and is higher than that from LST.
This could be due to the underlying assumptions in the $1D$-theory, and hints at a possible combination of modal and non-modal mechanisms during the disturbance evolution.
The wall-normal Fourier transformation of the temperature perturbation field at~$x=1.2$ corresponding to the waves in fig.~\ref{fig:2D_Amp}(a) are highlighted in fig.~\ref{fig:2D_Amp}(b).
The structure resembles that of modal entropic-instabilities in fig.~\ref{fig:Mode_LST}(b) with the peak closer to the boundary-layer edge.
The influence of $3D$-spatial inhomogeneities on the disturbance evolution is investigated next.

\subsection{Linear evolution of three-dimensional wavepackets}
\label{sec:3DWO}
\begin{figure}
\centering
\begin{subfigmatrix}{3}
\subfigure[]{{\includegraphics[trim=5 0 18 0,clip,width=0.325\textwidth]{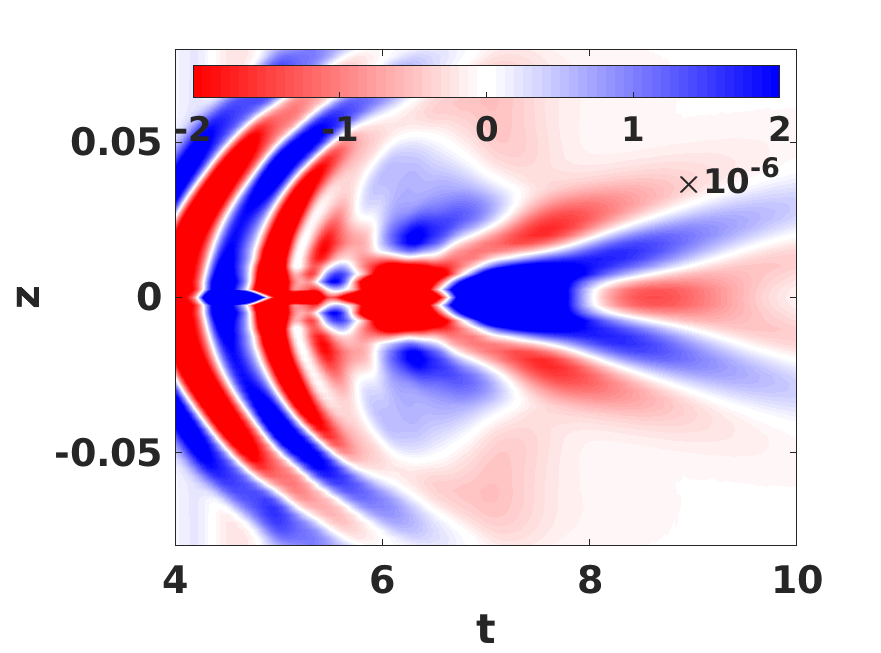}}}
\subfigure[]{{\includegraphics[trim=5 0 18 0,clip,width=0.325\textwidth]{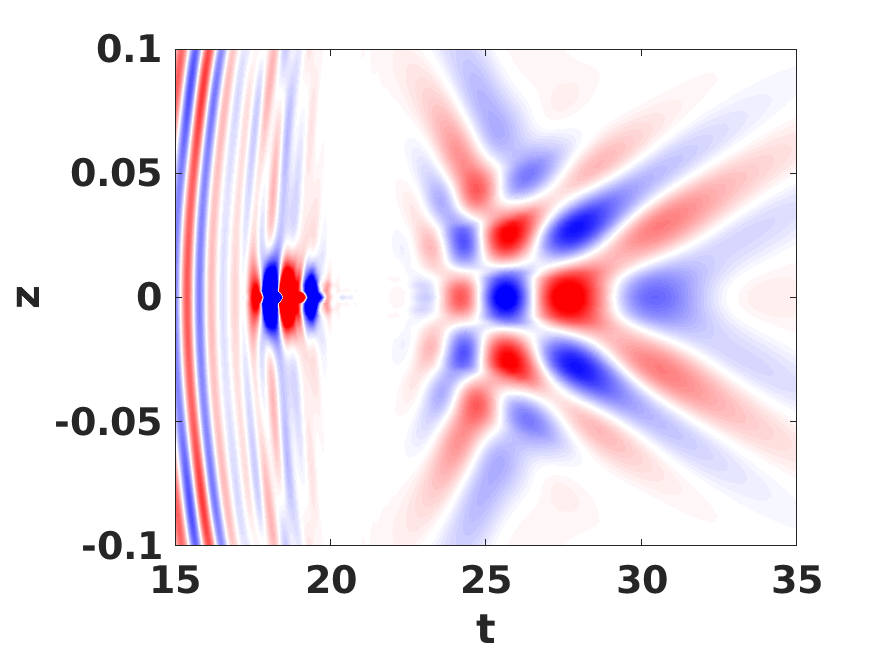}}}
\subfigure[]{{\includegraphics[trim=5 0 18 0,clip,width=0.325\textwidth]{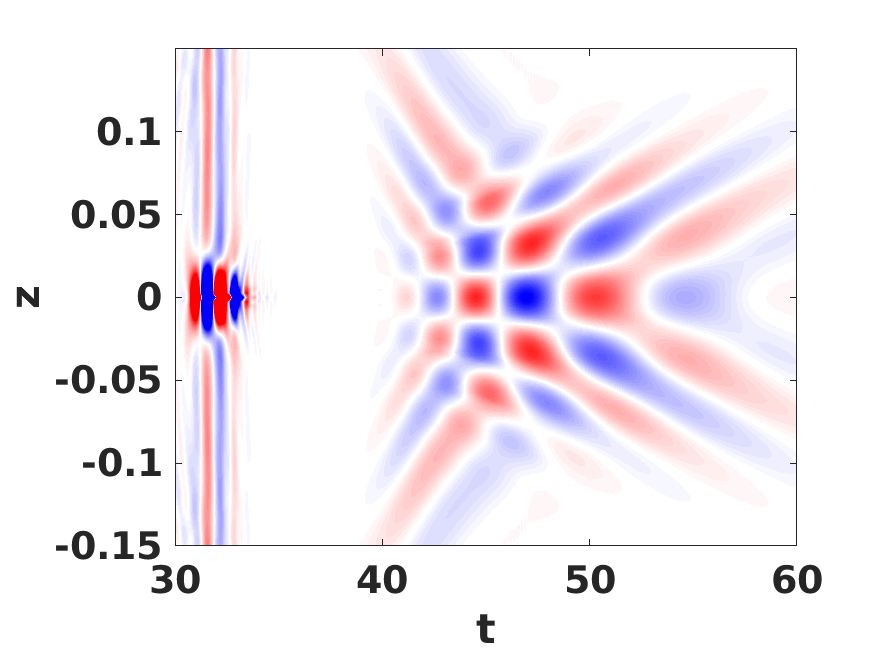}}}
\end{subfigmatrix}
\caption{Spanwise-temporal ($z-t$) pressure perturbation signature on the wall for forcing in the entropy-layer at $(a)~x=0.4$ $(b)~x=0.9$ $(c)~x=1.4$.}
\label{fig:EL_wpp}
\end{figure}

The development of three-dimensional perturbations is now examined by employing wavepacket analysis.
A short pulse, centered around~$St=20$, is excited in the entropy-layer, as outlined in Section~\ref{sec:methodology}.
The signature of wall pressure perturbation evolution from the wavepacket analysis is plotted in fig.~\ref{fig:EL_wpp}, with the $x$-axis adjusted to time scales.
Closer to the forcing location at~$x=0.4$ in fig.~\ref{fig:EL_wpp}(a), somewhat semicircular (acoustic) waves emanate ($t\in(4,6)$) and are likely associated with the suddenly imposed pulse forcing.
These waves travel faster than the wavepacket~(see~$t\sim15$ in fig.~\ref{fig:EL_wpp}(b) at~$x=0.9$) and do not amplify. 
Quasi-$2D$ waves trail these semicircular wave-front at~$t\sim20$, followed by oblique~$3D$ disturbances at~$t\sim 25$.
A lower frequency of these oblique waves than the leading $2D$~disturbances is apparent.
In fig.~\ref{fig:EL_wpp}(c) at~$x=1.4$, oblique disturbances spread spanwise and evidently convect at lower-velocities than the leading $2D$~waves.
Symmetric development of the disturbances is observed in the spanwise direction.
\new{Additional linear wavepacket simulations centered around $St=20$ were performed with a smooth monopole forcing function in the spanwise extent (along with wall-normal and streamwise variations) -- the disturbance evolution features remained the same.}

\begin{figure}
\centering
\begin{subfigmatrix}{2}
\subfigure[]{{\includegraphics[trim=5 0 18 0,clip,width=0.325\textwidth]{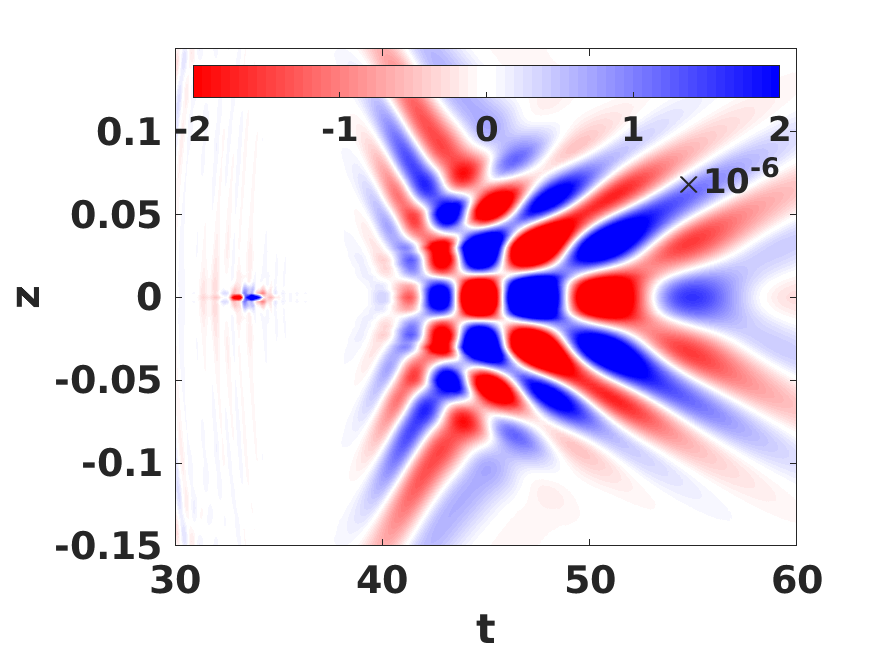}}}
\subfigure[]{{\includegraphics[trim=5 0 18 0,clip,width=0.325\textwidth]{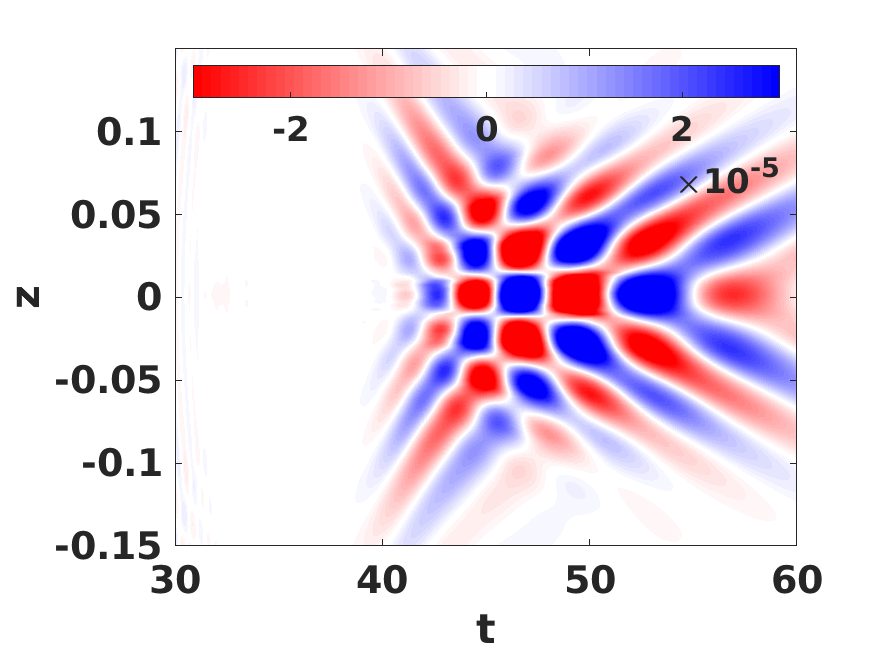}}}
\end{subfigmatrix}
\caption{ Spanwise-temporal ($z-t$) pressure perturbation signature on the wall at~$x=1.4$ for $(a)$~Boundary-layer edge forcing $(b)$~Wall forcing.}
\label{fig:WPP}
\end{figure}

To examine the role of the actuator location on the disturbance evolution, complimentary wavepacket analyses were performed through forcing at the boundary-layer edge and the wall.
The wall pressure perturbation signatures for these cases are highlighted in fig.~\ref{fig:WPP}(a,b), respectively at $x=1.4$.
While the features of oblique waves are similar in the forcing conditions considered, the leading $2D$~waves are very weak in fig.~\ref{fig:WPP}(b,c), compared to the the wavepacket initiated in the entropy-layer.
This can be attributed to the diminished receptivity of the entropy-layer waves, consistent with the previous $2D$~calculations.
The oblique waves are the signature of the boundary-layer instabilities and are efficiently triggered by forcing at all wall-normal locations considered.
The above simulations were repeated for multiple wavepackets centered at $St=15$,~$25$, and the conclusions remain unaltered.

\begin{figure}
\centering
\begin{subfigmatrix}{3}
\subfigure[]{{\includegraphics[trim=0 0 0 0,clip,width=0.325\textwidth]{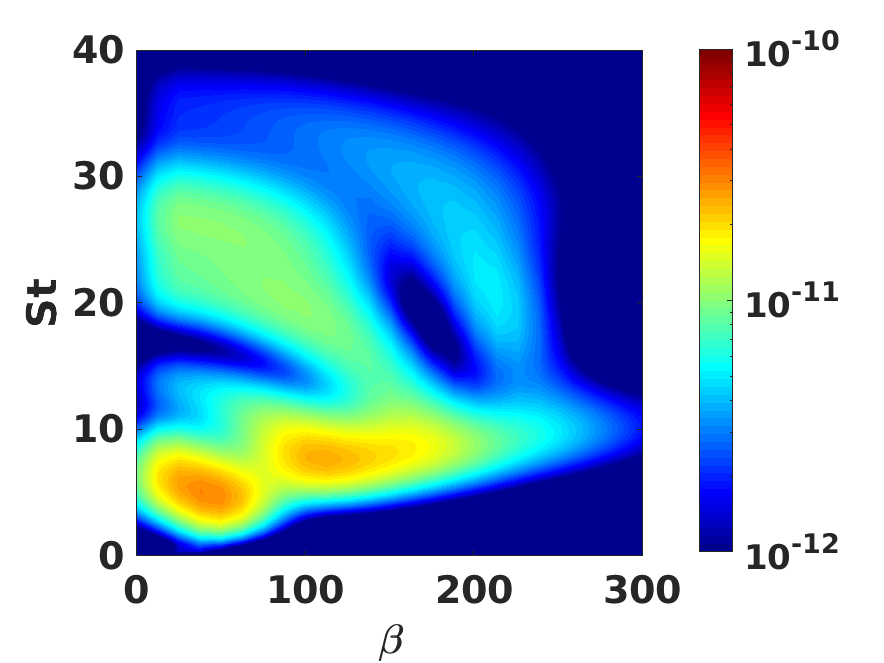}}}
\subfigure[]{{\includegraphics[trim=0 0 0 0,clip,width=0.325\textwidth]{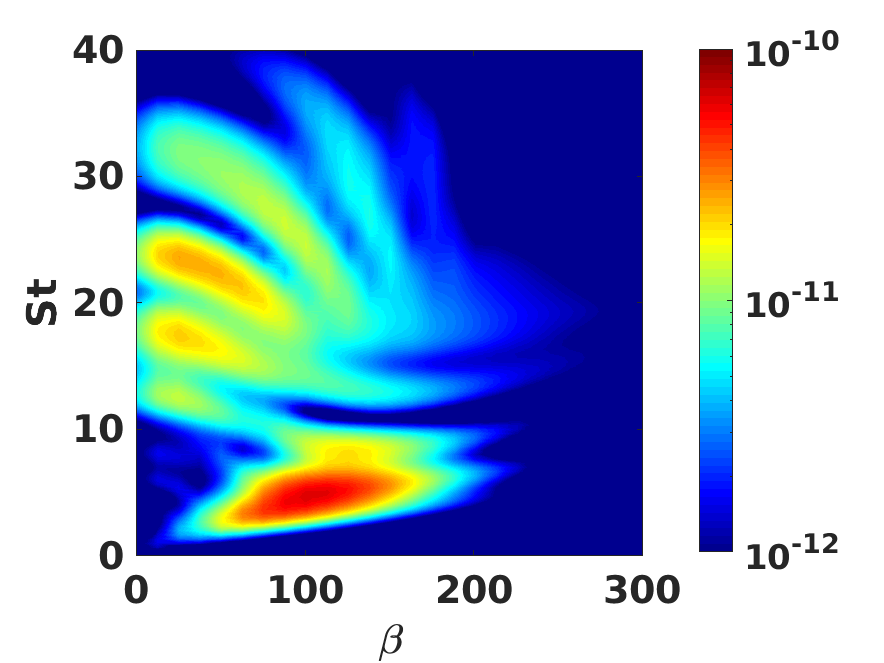}}}
\subfigure[]{{\includegraphics[trim=0 0 0 0,clip,width=0.325\textwidth]{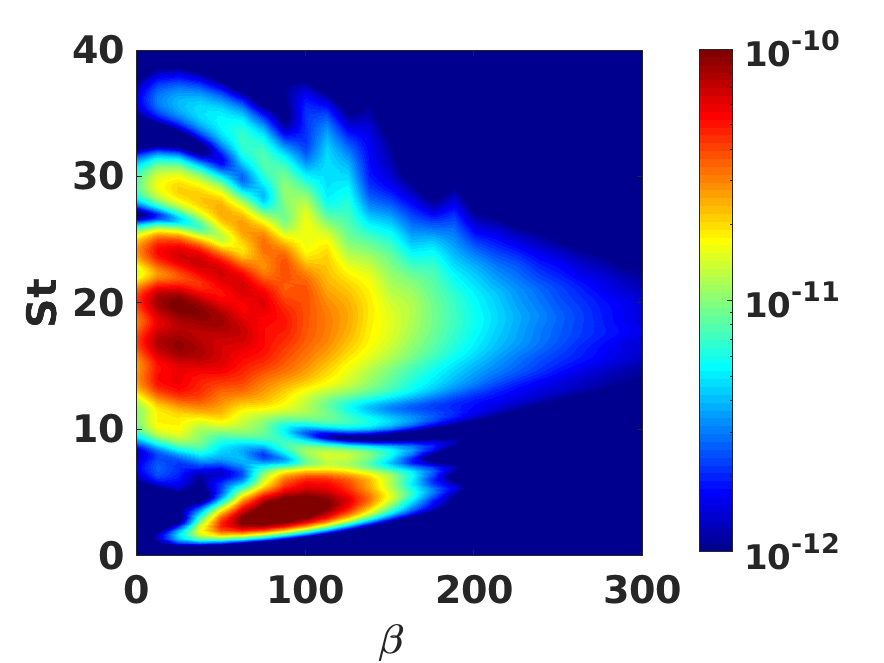}}}
\end{subfigmatrix}
\caption{Pressure perturbation spectral signature in the frequency-spanwise wavenumber~$(St,\beta)$ plane on the wall for forcing in the entropy-layer at the streamwise locations $(a)~x=0.4$ $(b)~x=0.9$ $(c)~x=1.4$.}
\label{fig:EL_fft}
\end{figure}

The spectral content of the wall pressure perturbations are examined in fig.~\ref{fig:EL_wpp} by employing two-dimensional Fourier transformations.
The corresponding frequency-spanwise wavenumber $(St,\beta)$ spectra and (logarithmic) amplitude of the disturbances are highlighted in fig.~\ref{fig:EL_fft}.
Broadband spectral content is observed closer to the forcing location at $x=0.4$ in fig.~\ref{fig:EL_fft}(a).
Amplification of low- $(St=5)$ and high- $(St=20)$ frequency waves at $x=0.9$ can be inferred from fig.~\ref{fig:EL_fft}(b).  
Further downstream at $x=1.4$ in fig.~\ref{fig:EL_fft}(c), the wavepacket exhibits amplification of low-frequency waves around $St=5$.
Oblique waves also exhibit growth in the high-frequency range around $St=20$.
While the lower-frequency waves correspond to the oblique waves discussed in fig.~\ref{fig:EL_wpp}, the higher-frequency waves are the signature of the leading quasi-$2D$ waves.
Consistent results were obtained from the spectral analysis of the pressure perturbations from the forcing at boundary-layer edge, wall, and are not shown here. 

\begin{figure}
\centering
\begin{subfigmatrix}{2}
\subfigure[]{{\includegraphics[width=0.49\textwidth]{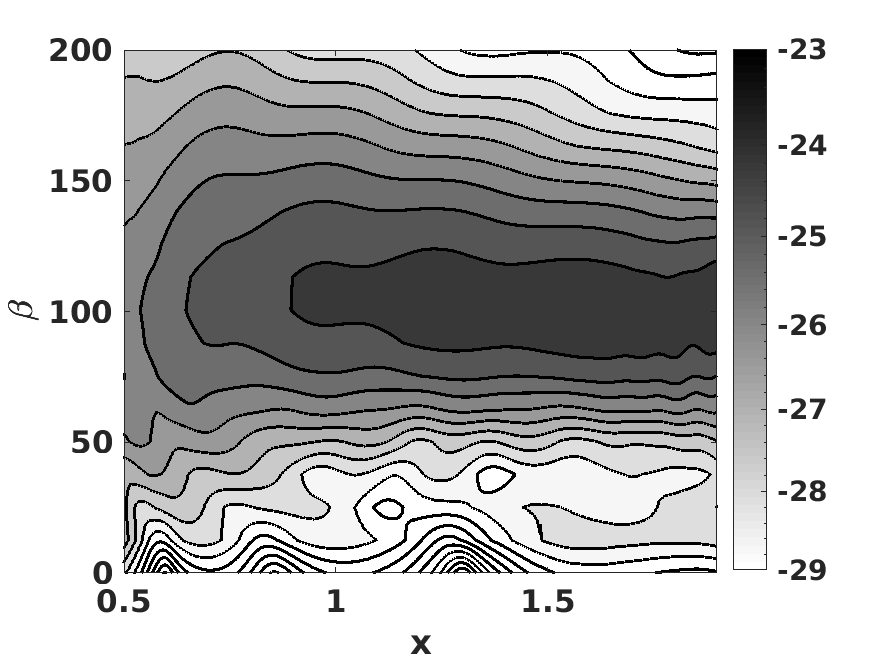}}}
\subfigure[]{{\includegraphics[width=0.49\textwidth]{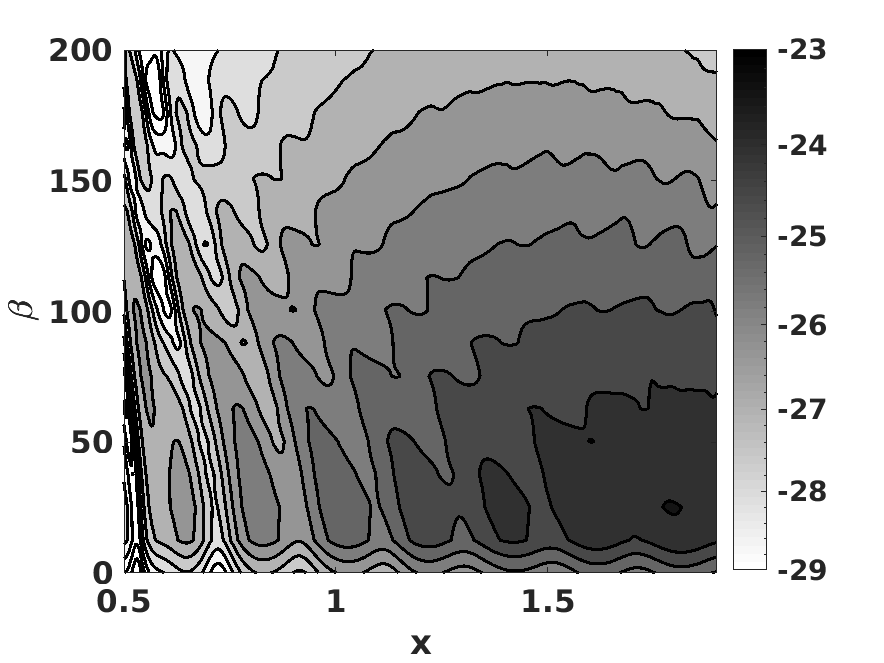}}}
\end{subfigmatrix}
\caption{Pressure perturbation (logarithmic) magnitude contours for the streamwise variation of spanwise wavenumber $(\beta-x)$, with forcing in the entropy-layer.
The response of the wavepacket at the frequencies $(a)~St=5$ $(b)~St=20$.}
\label{fig:WP_Stx}
\end{figure}

Further distinctions between the evolution features of instabilities generated in the boundary- and entropy-layers can be made from the streamwise-wavenumber contours of the pertinent frequencies.
Figure~\ref{fig:WP_Stx}(a) shows the development of the lower-frequency boundary-layer waves at $St=5$.  
Larger amplification of (first-mode) oblique waves around $\beta=100$ is evident, suggesting the dominance of boundary-layer instabilities over the entropic-instabilities at lower frequencies.
The contours trace a typical \textit{thumb curve} in the streamwise-spanwise plane, resembling the first-mode instability.
Entropic disturbances display amplification signatures over a wider spanwise extent, planar and oblique, as observed in fig.~\ref{fig:WP_Stx}(b).
These contours do not resemble typical first- or second-mode neutral stability curves.
\new{Note that the $t^{-1/2}$ dependence of amplitude due to wavepacket spreading~\cite{gallaire2003mode} has not been considered here based on the earlier works of Fasel and co-workers~\cite{sivasubramanian2014numerical, hader2019direct, hartman2021nonlinear} on similar configurations.}

\new{The role of these disturbances in turbulence onset is discussed next.
In free-flight or wind tunnels, the receptivity process depends on the freestream spectra, which in turn dictate the nature of instability evolution.
Due to lack of complete information about these disturbances (frequencies, wavenumbers, and amplitudes), it is challenging to accurately incorporate them into transition prediction models~\cite{schneider2004hypersonic}.
Another approach to study transition is to isolate dominant disturbance features and examine their evolution.
Such \textit{controlled} breakdown simulations are inspired by the vibrating ribbon experiments by Schubauer and Skramstad~\cite{schubauer1947laminar} in low-speed flows.
Although controlled breakdown experiments have not yet been performed on blunted objects, they could be realized in future high-speed experiments~\cite{hartman2021nonlinear}.}

Based on the symmetric development of the wavepacket in fig.~\ref{fig:EL_wpp} and the characteristic streamwise disturbance amplification, oblique breakdown of both boundary- and entropy-layer instabilities are potential transition mechanisms.
\new{Other mechanisms involving multi-modal interactions between boundary- and entropic-instabilities may also lead to transition, and are not pursued in this paper.
The focus is on the transition induced by oblique disturbances in the entropy-layer and qualitatively comparing the breakdown mechanisms with blunted cones.
}

\subsection{Nonlinear evolution of oblique entropic disturbances}
\label{sec:3DDNS}
\subsubsection{Coherent structures during oblique breakdown}

To study the laminar-turbulent transition, two oblique entropic disturbances are continuously excited in the entropy-layer.
The frequency~$St=20$ and spanwise wavenumbers~$\beta=\pm 70$ are chosen to highlight the processes occurring during transition.
Note that at~$St=5$, these spanwise wavenumbers also exhibit boundary-layer instability amplification, allowing, if desired, the characterization of oblique breakdown including entropy-layer effects.
To focus computational resources in the wave amplification regime, the inflow of the domain is placed at~$x=0.5$ and the exit is at $x=1.95$. 
The spanwise domain is $z \in [\pm\pi/70] \sim [\pm 0.0448]$, and is resolved by~$128$ points.

\begin{figure}
\centering
\begin{subfigmatrix}{2}
\subfigure[]{{\includegraphics[width=0.9\textwidth]{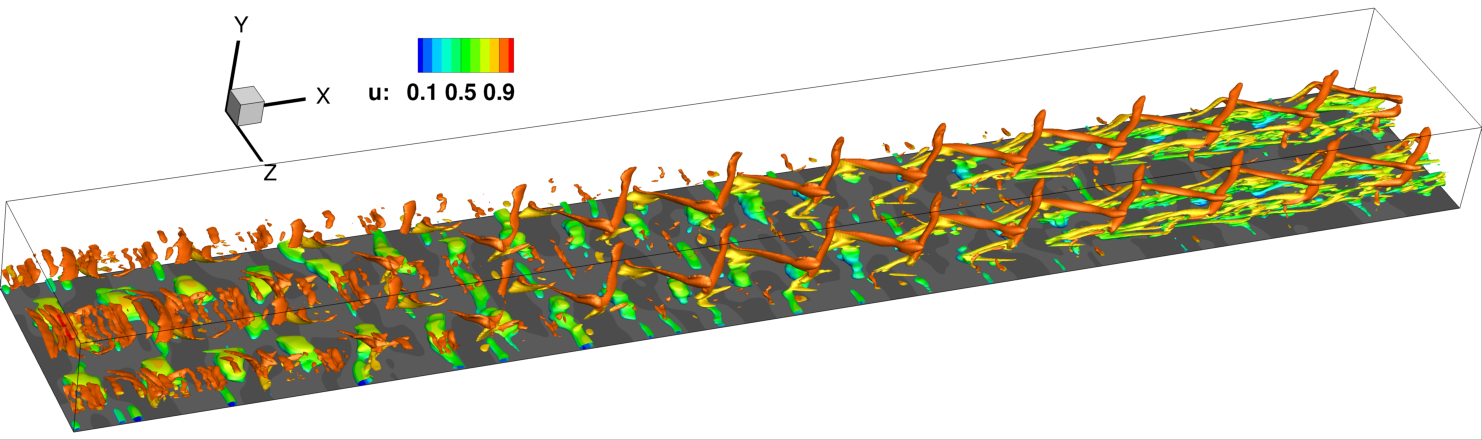}}}
\subfigure[]{{\includegraphics[width=0.9\textwidth]{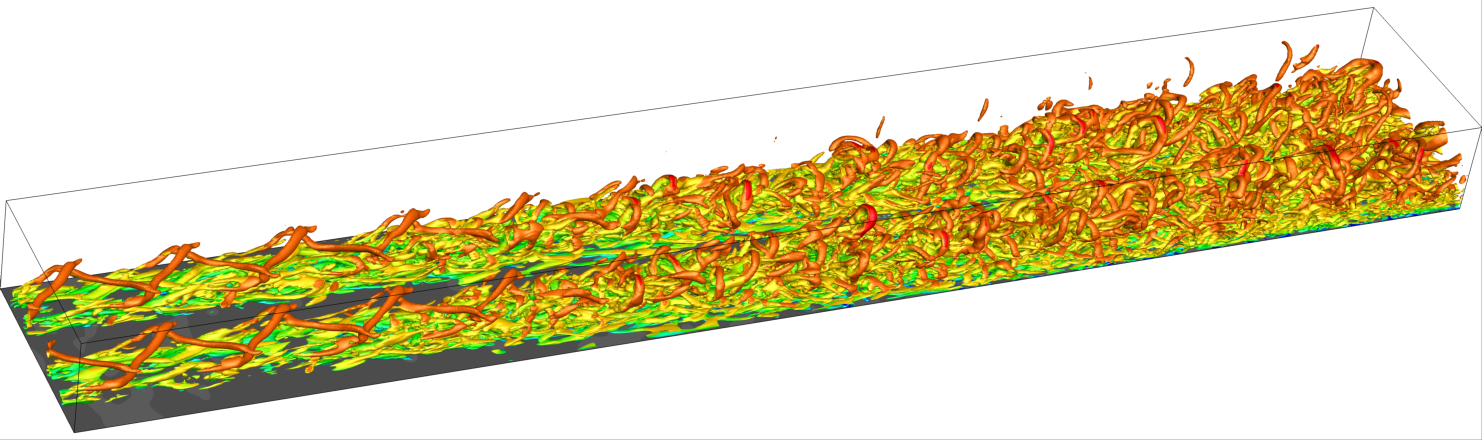}}}
\end{subfigmatrix}
\caption{Vortical structures during oblique breakdown identified by Q-criterion iso-surfaces~$(Q~=~100)$, colored by streamwise velocity~$u$.
Each frame represents equal increments in the streamwise extent as: $(a)$~$0.7<x<1.3$ $(b)$~$1.3<x<1.9$.}
\label{fig:qcrit}
\end{figure}

The nonlinear evolution of the above flow field, identified by Q-criterion iso-surfaces $(Q~=~100)$ and colored by streamwise velocity, is depicted in fig.~\ref{fig:qcrit}(a) for $0.7\leq x \leq 1.3$ and fig.~\ref{fig:qcrit}(b) for $1.3\leq x \leq 1.9$.
The Q-criterion, $Q=0.5[|\Omega|^2-|\textbf{S}|^2]$ uses $\Omega$~and~$\textbf{S}$ as antisymmetric and symmetric components of velocity-gradient~\cite{jcr1988eddies}, to aid in identifying dominant vortical structures.
In fig.~\ref{fig:qcrit}(a), vortical structures generated due to the interaction of high-amplitude waves at the forcing location convect and decay in the streamwise direction.
While no significant amplification of near-wall coherent structures is observed, a zigzag shaped organization arises in the central part of the panel in the entropy-layer.
At the end of the subdomain, near-wall streaks and the zigzag pattern can be more clearly visualized to be an overlap of two spanwise inclined structures. 
In the incompressible regime, Brandt and Henningson~\citep{brandt2002transition} highlighted such a staggered arrangement of vortices in the sinuous breakdown of boundary-layer streaks on a flat plate.  
These were also reported during the oblique breakdown on hypersonic blunt cones by Paredes et al.~\cite{paredes2020mechanism}, Hartman et al.~\cite{hartman2021nonlinear}.  
These structures further intensify, leading to the formation of a \textit{lambda} pattern evident at the beginning of the subdomain in fig.~\ref{fig:qcrit}(b).
\new{Late nonlinear stages of both K-type and oblique breakdown exhibit aligned \textit{lambda} vortices.
K-type transition requires two low-amplitude oblique waves along with a high-amplitude planar wave (at same temporal frequencies) in the freestream or inlet spectra~\cite{hader2019direct}.
These conditions are not satisfied by the current controlled forcing or ensuing nonlinear interactions.
Hence, following the discussion of Berlin et al.~\cite{berlin1999numerical}, the structures observed here do not correspond to K-type transition.}
Closer to the wall, the streaks grow in the streamwise and spanwise directions below the zigzag arrangement.
The vortical structures further fill the spanwise domain; the signature of the staggered arrangement cannot be discerned at the downstream end of the domain. 
The fine-scale structures and the preponderance of hairpin vortices at the end of the subdomain are characteristic of early turbulent boundary-layers, as suggested by Eitel-Amor et al.~\cite{eitel2015hairpin}.

\begin{figure}
\centering
\begin{subfigmatrix}{3}
\subfigure[]{{\includegraphics[trim=10 80 30 90,clip,width=0.6\textwidth]{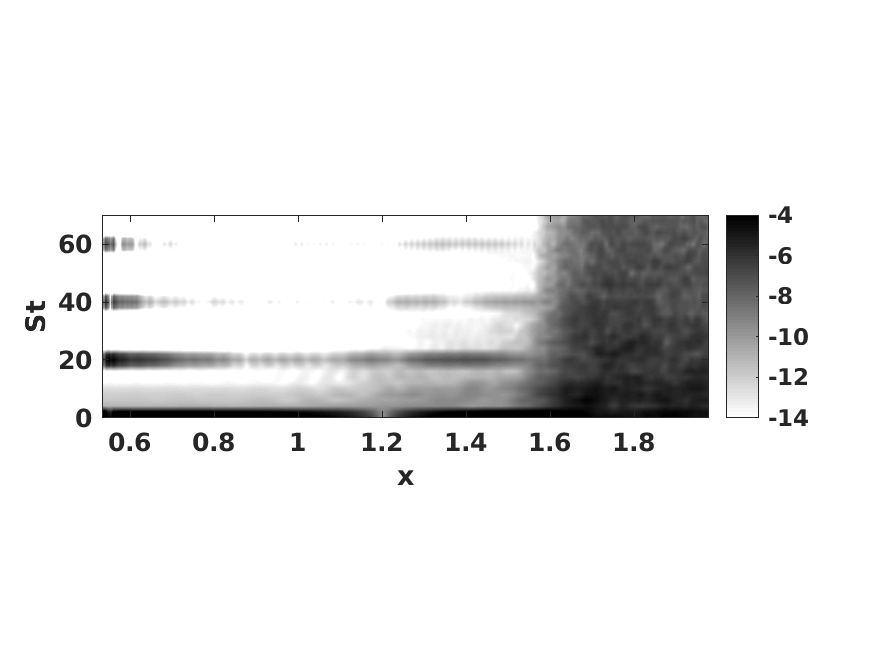}}}
\subfigure[]{{\includegraphics[trim=0 50 40 70,clip,width=0.49\textwidth]{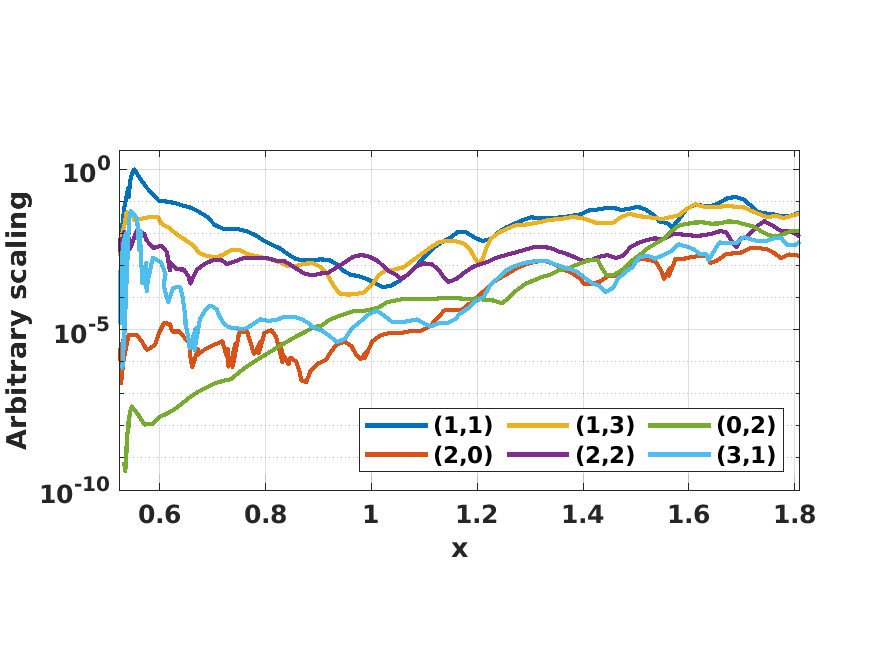}}}
\subfigure[]{{\includegraphics[trim=0 50 40 70,clip,width=0.49\textwidth]{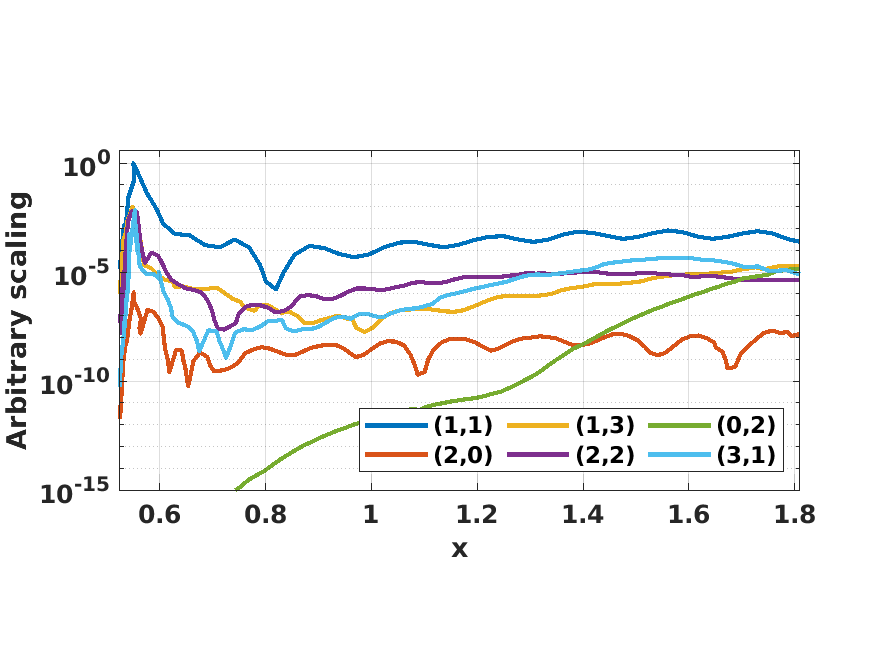}}}
\end{subfigmatrix}
\caption{Spectral features of the transitional flow.
$(a)$ Frequency~$(St)$ distribution of the perturbations~($u'$) in the streamwise extent~$(x)$ at~$z=0$.
Evolution of \textit{signature-modes} close to the wall for $(b)$ high-amplitude $(c)$ low-amplitude forcing conditions.}
\label{fig:Stx}
\end{figure}

A quantitative description of the spectral filling during the transition is provided in fig.~\ref{fig:Stx} using near wall~$(y\sim 2\times 10^{-4})$ velocity perturbation~$(u')$ data.
The streamwise evolution of the frequency spectrum ~$(St-x)$ (in logarithmic scale) at the center-(symmetry) plane~$(z=0.0)$ is highlighted in fig.~\ref{fig:Stx}(a).
Near the forcing location~$(x \sim 0.55)$, harmonics of the forcing waves ~$(St=20)$ are observed at $St=40$ and $60$, along with low-frequency waves near $St \sim 0$.
The amplitude of the forced disturbances initially decreases until~$x\sim 0.8$, but is followed by an increase until~$x\sim 1.6$.
The higher harmonics decay rapidly near the forcing location and can be associated with the near-wall signature of the damping entropy-layer structures described in fig.~\ref{fig:qcrit}(a). 
The signature of low-frequency waves at $St \sim 0$ persists until $x\sim 1.6$.
In the range~$1.1 < x <1.6$, gradual spectral broadening is observed, and for $x > 1.6$ the frequency spectrum is fully populated.
The former range corresponds to the instability development shown earlier in fig.~\ref{fig:qcrit}(a), and the latter suggests late stages of transition in fig.~\ref{fig:qcrit}(b). 

\new{The spectral features are further highlighted using two-dimensional Fourier transformations to the near-wall streamwise velocity perturbations (in $z-t$ plane).}
The streamwise evolution of dominant frequency-wavenumber $(St,\beta)$ pairs are plotted in fig.~\ref{fig:Stx}(b), with a shorthand notation~$(n,m)$ \cite{hader2019direct}, where $n$~indicates the multiples of forcing frequency~$(St=20)$ and $m$~indicates multiples of the spanwise wavenumber~$(\beta=70)$.
Here, the evolution of so-called \textit{signature-modes} of oblique breakdown~\cite{fasel1993direct} are depicted.
The forced wave~$(1,1)$ decays initially followed by an exponential growth in~$1.0 < x < 1.4$.
The interaction of~$(1,\pm1)$ generates a vortex~$(0,2)$, which is amplified by the lift-up mechanism~\cite{landahl1975wave}, resulting in streamwise streaks.
These grow in amplitude downstream by a continuously forced transient growth mechanism~\cite{laible2016continuously}.
Other modes considered here initially decay followed by an exponential growth till~$x\sim 1.5$.
The above exponentially growing waves (nearly) saturate at~$x \sim 1.6$, exchanging energy across different scales, resulting in spectral filling as shown fig.~\ref{fig:Stx}(a) and multi-scale structures in fig.~\ref{fig:qcrit}(b). 
In the blunt cone study by Hartman et al.~\cite{hartman2021nonlinear}, besides an initial decay followed by growth, a second reduction in the modal amplification before breakdown was reported.
The latter behavior is not observed in the current scenario, and after the initial growth, the flow transitions to turbulence.

\new{Additional lower-amplitude simulations have been performed by decreasing the inlet wave amplitudes by~$10$ times and the spectral signature is shown in fig.~\ref{fig:Stx}(c).
The evolution of the signature-modes suggests streak amplification is the largest, similar to higher amplitude case.}
Other signature modes,~$(1,3)$,~$(3,1)$, and $(2,2)$ also exhibit commensurate amplification trends in both cases.  
A notable difference is in the evolution of the~$(2,0)$ mode which does not exhibit significant amplification in the low-amplitude simulations.
\new{Additionally, the streak mode equilibrates earlier with other modes in higher amplitude case.
These differences can be attributed to the streak strength, enhanced nonlinear interactions in high amplitude case and will be elucidated in later sections.}

\begin{figure}
    \centering
    {{\includegraphics[trim=10 160 40 200,clip,width=0.9\textwidth]{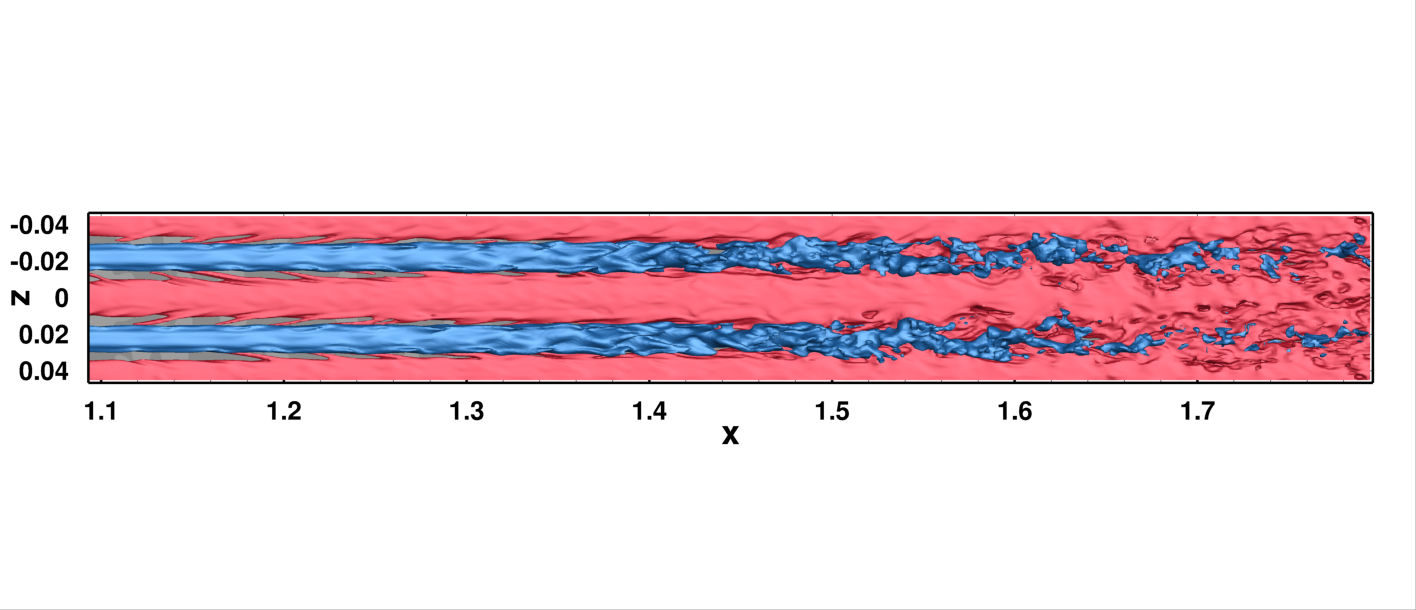}}}
    \caption{Top view of the instantaneous streamwise velocity perturbations.
    Blue and red iso-surfaces represent positive and negative fluctuations respectively, with a magnitude $|u'|=0.15$.}
    \label{fig:up_iso}
\end{figure}

To visualize the breakdown process, an instantaneous top view of streamwise velocity  fluctuations in the range~$1.1 < x< 1.8$ is plotted in fig.~\ref{fig:up_iso}.
Here, the fluctuation is defined as the deviation of the transitional flow field from the laminar base flow, and the blue and red isosurfaces indicate negative and positive values with an amplitude~$|u'|=0.15$.
The negative velocity isosurface indicates streak-like structure centered around~$x \sim \pm 0.02$.
In earlier supersonic oblique transition studies~\cite{franko2013breakdown, mayer2011direct, fasel1993direct} alternating low-momentum/high-temperature and high-momentum/low-temperature streaks have been reported. 
The initially horizontal structure $(x < 1.12)$ exhibits spanwise asymmetric oscillations (\textit{S}- shaped) in the range $1.14 < x< 1.2$, indicating the onset of streak secondary instability.
These oscillations further intensify in~$1.35< x< 1.5$ followed by the breakdown of the low-speed streak for~$x> 1.6$.
On the flanks of the positive velocity streaks, spanwise inclined symmetric structures can be observed at $x \sim 1.14$, ~$1.18$ and $z \sim \pm 0.01$.
The high-speed streak loses its coherence downstream, indicating its breakdown.
The (anti)symmetric oscillation of (low)high-speed streaks suggests the subharmonic sinuous fluctuations~\cite{andersson2001breakdown} to be the dominant secondary instability of the streaks.

\begin{figure}
\centering
\begin{subfigmatrix}{3}
\subfigure[]{{\includegraphics[trim=5 0 18 0,clip,width=0.325\textwidth]{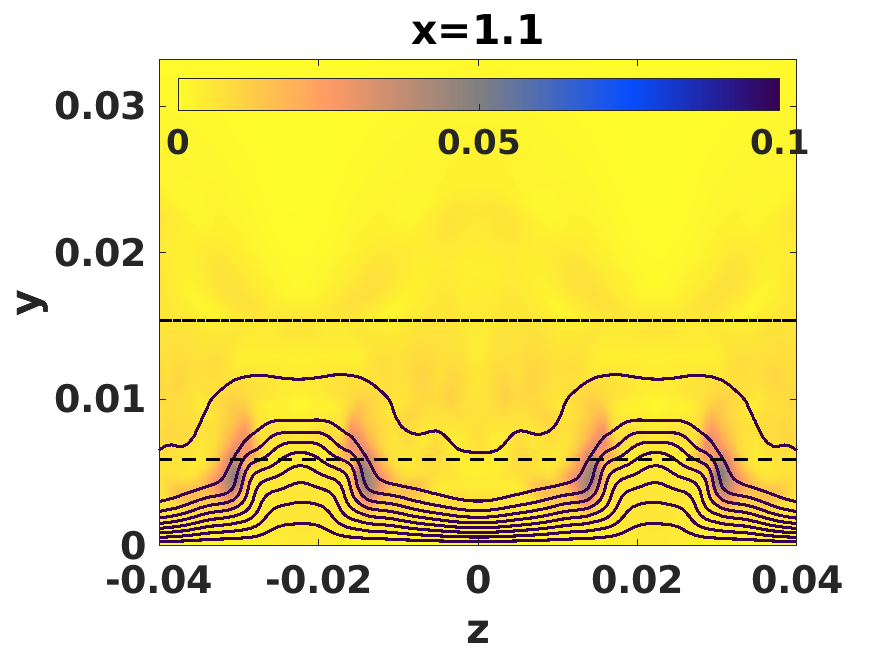}}}
\subfigure[]{{\includegraphics[trim=5 0 18 0,clip,width=0.325\textwidth]{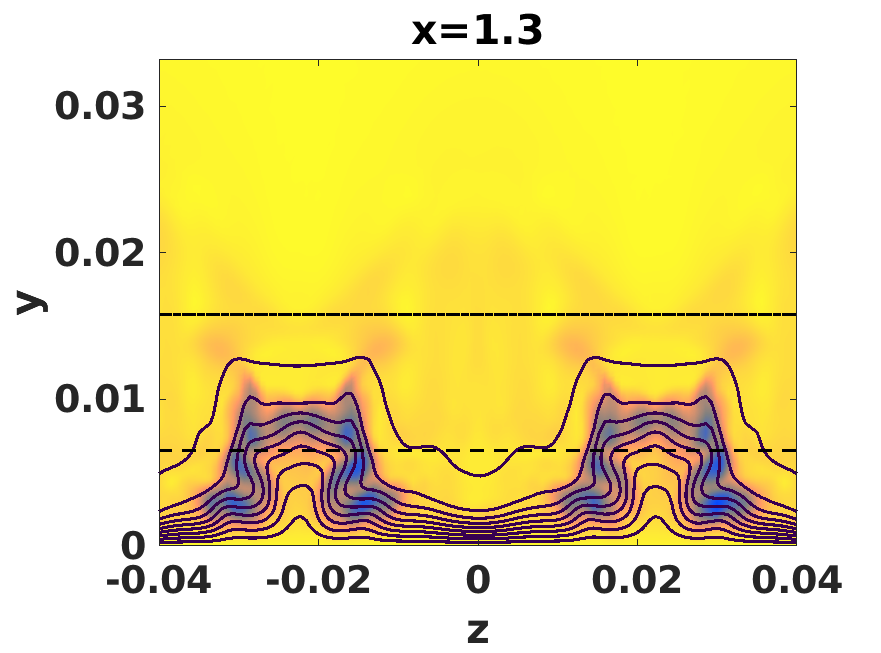}}}
\subfigure[]{{\includegraphics[trim=5 0 18 0,clip,width=0.325\textwidth]{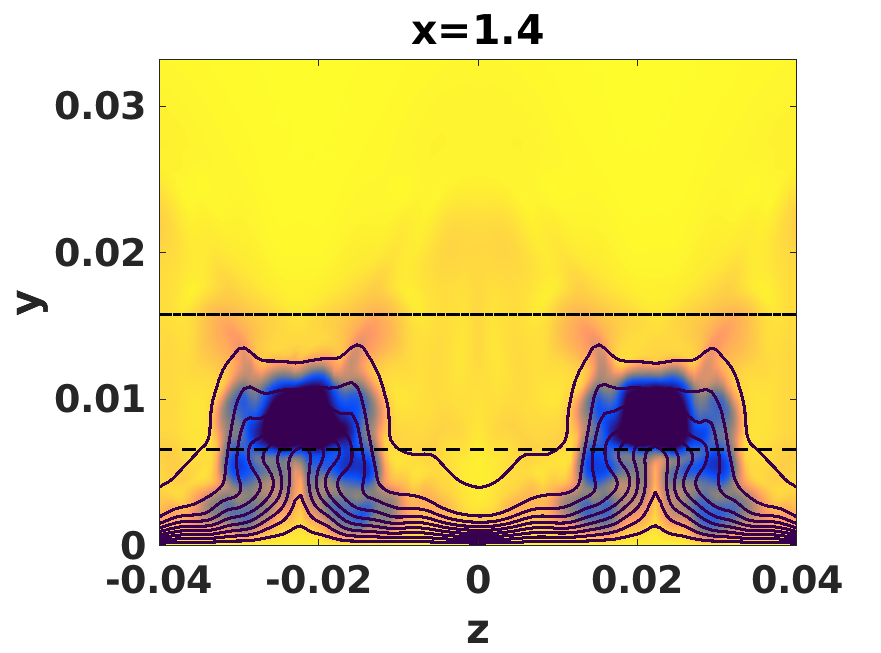}}}
\subfigure[]{{\includegraphics[trim=5 0 18 0,clip,width=0.325\textwidth]{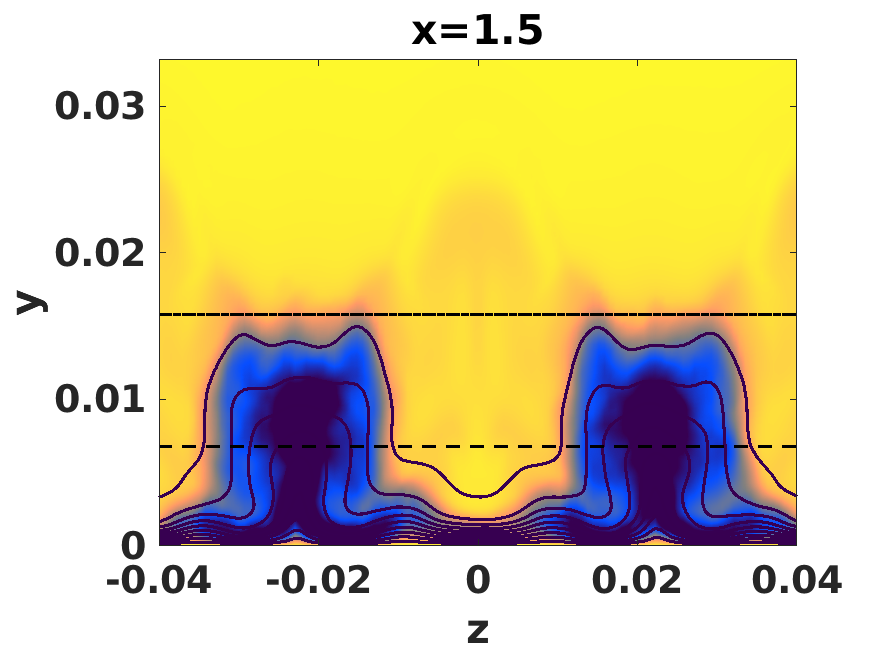}}}
\subfigure[]{{\includegraphics[trim=5 0 18 0,clip,width=0.325\textwidth]{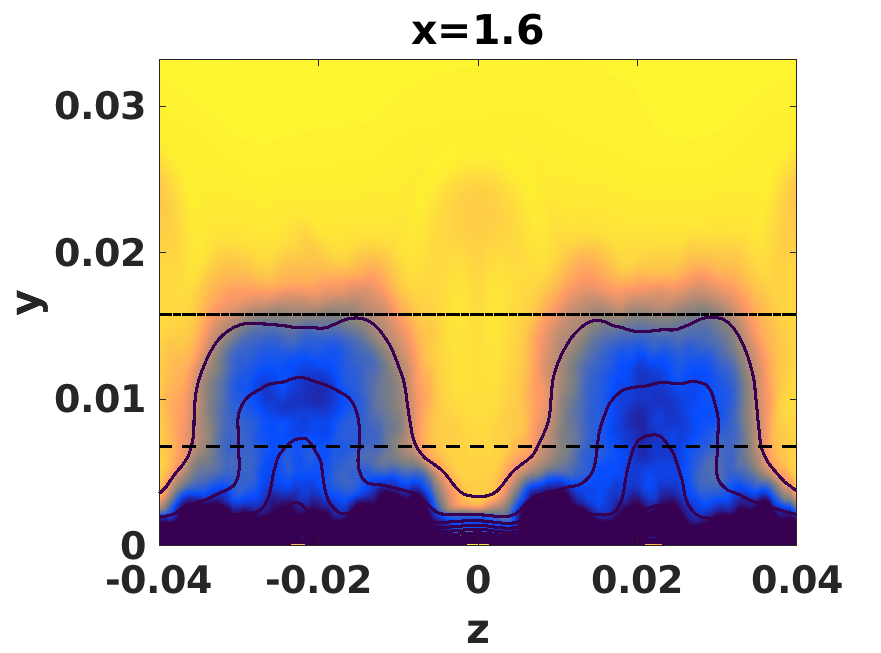}}}
\subfigure[]{{\includegraphics[trim=5 0 18 0,clip,width=0.325\textwidth]{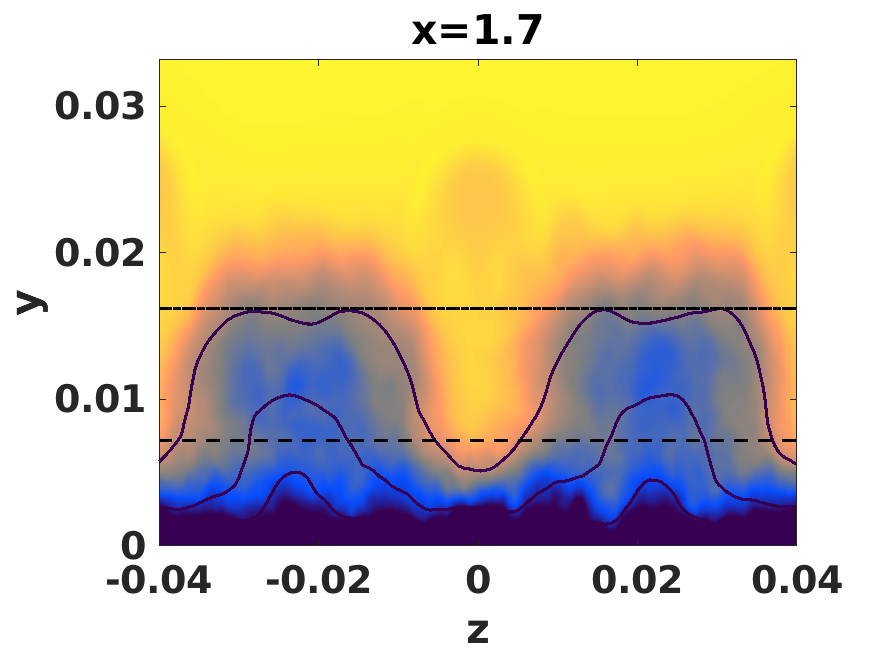}}}
\end{subfigmatrix}
\caption{Evolution contours (colored) of streamwise velocity root mean square (rms) fluctuations.
Contour lines represent time-averaged streamwise velocity.
Entropy- and boundary layer-edges are marked by dotted-dashed~$(.-)$ and dashed~$(--)$ lines, respectively.
\new{Colorbars are same for each subplot and are shown in panel~$(a)$}.}
\label{fig:urms}
\end{figure}

The dynamics of the disturbance evolution in the cross-flow planes~$(y,z)$ are now examined by considering the root mean square (rms) amplitude evolution of the streamwise velocity fluctuations after the initial transients have exited the domain.
In fig.~\ref{fig:urms}, contours of rms velocity at different streamwise locations (filled colors, \new{same contour levels}) and overlaid with time-averaged streamwise velocity contours (lines)
The panels contain ten equally spaced mean velocity contour lines (from~$0.1$ to $1.0$) while the fluctuation color scale ranges from~$0$ to $0.1$.
In fig.~\ref{fig:urms}(a), the mean velocity contours indicate the penetration of low-speed streaks into the entropy-layer.
Fluctuations are observed in the negative streaks at various wall normal locations~($y \sim 0.0025$,~$0.005$,~$0.01$), both in the boundary- and the entropy-layers.
Note that these fluctuations are dominant where the mean flow profile exhibits \textit{kinks}, which are inflection points that can support inviscid secondary instabilities~\cite{andersson2001breakdown}.
Further intensification of these fluctuations is observed in fig.~\ref{fig:urms}(b) and disturbances are evident between high- and low-speed streaks closer to the wall.  
In fig.~\ref{fig:urms}(c) perturbations are strengthened closer to the crest; their signatures are apparent in fig.~\ref{fig:up_iso} as \textit{warping} of the low-speed streak isosurface.    
In blunt cone transition studies, Paredes et al.~\cite{paredes2020mechanism} reported high magnitude perturbations azimuthally closer to the crest of the streaks as in fig.~\ref{fig:urms}(c), but not the near-wall signature in fig.~\ref{fig:urms}(a,b). 
Vaughan and Zaki~\cite{vaughan2011stability} demonstrated two types of streak destabilization in incompressible flows, namely \textit{outer} and \textit{inner} instabilities, based on the height of their respective critical layers from the wall. 
The former is related to the secondary instability of lifted low-speed streaks subject to high-frequency freestream forcing, while the latter reside nearer to the wall.
Here, the boundary-layer disturbances in fig.~\ref{fig:urms}(b) bear resemblance to the inner instabilities, while the fluctuations in fig.~\ref{fig:urms}(c) suggest outer streak instabilities.
However, in the limit of zero streak amplitude, the near-wall fluctuations alone are not supported at $St=20$ (similar to the argument in fig.~\ref{fig:rhop}), suggesting that they are not inner instabilities.
This signature could be a result of the additional shear layers in the current forcing configuration, distinct from that of the optimal non-modal inlet forcing on blunt cones.
Perturbation energy transfers from the crest of streaks (in the entropy-layer) to the valleys (in the boundary-layer) as noticed in fig.~\ref{fig:urms}(d).
At~$x=1.6$, in fig.~\ref{fig:urms}(e), reduction in the amplitude at the crest and further increase in the near-wall fluctuations are observed.
In fig.~\ref{fig:urms}(f), a spanwise homogenization trend relative to the preceding location indicates the late nonlinear stages of transition.

\begin{figure}
\centering
\begin{subfigmatrix}{1}
\subfigure[]{{\includegraphics[trim=0 110 10 120,clip,width=1.0\textwidth]{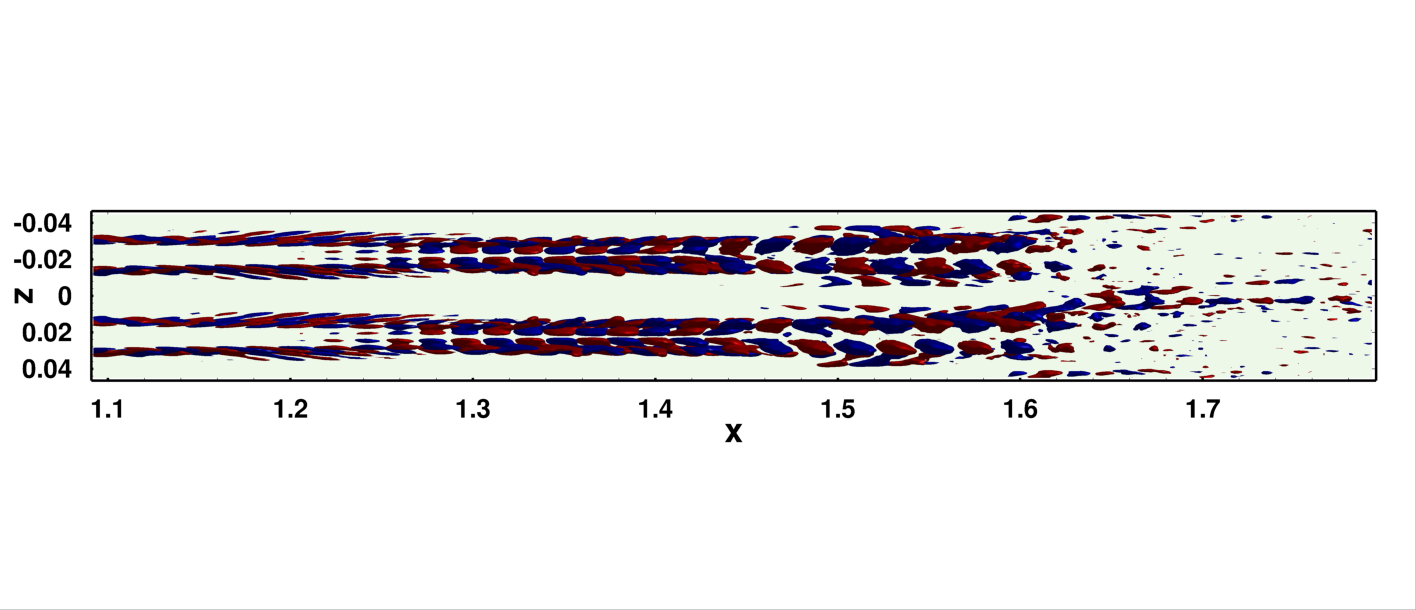}}}
\subfigure[]{{\includegraphics[trim=0 110 10 120,clip,width=1.0\textwidth]{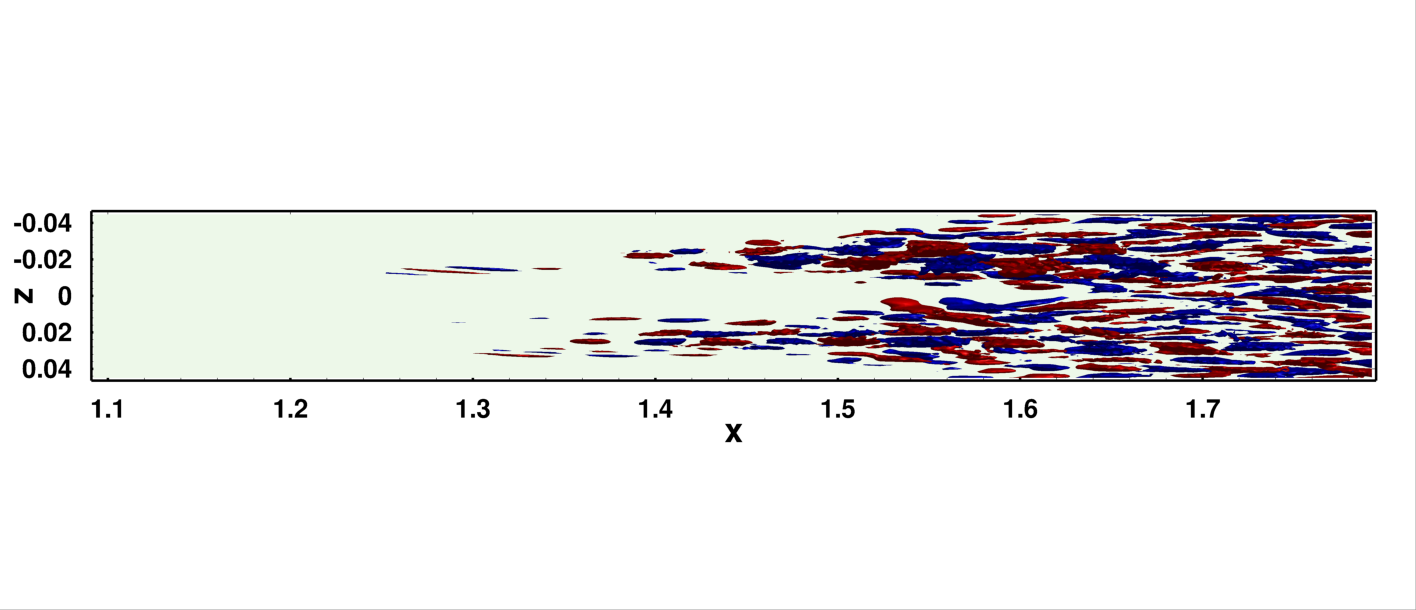}}}
\end{subfigmatrix}
\begin{subfigmatrix}{3}
\subfigure[]{{\includegraphics[trim=5 0 18 0,clip,width=0.325\textwidth]{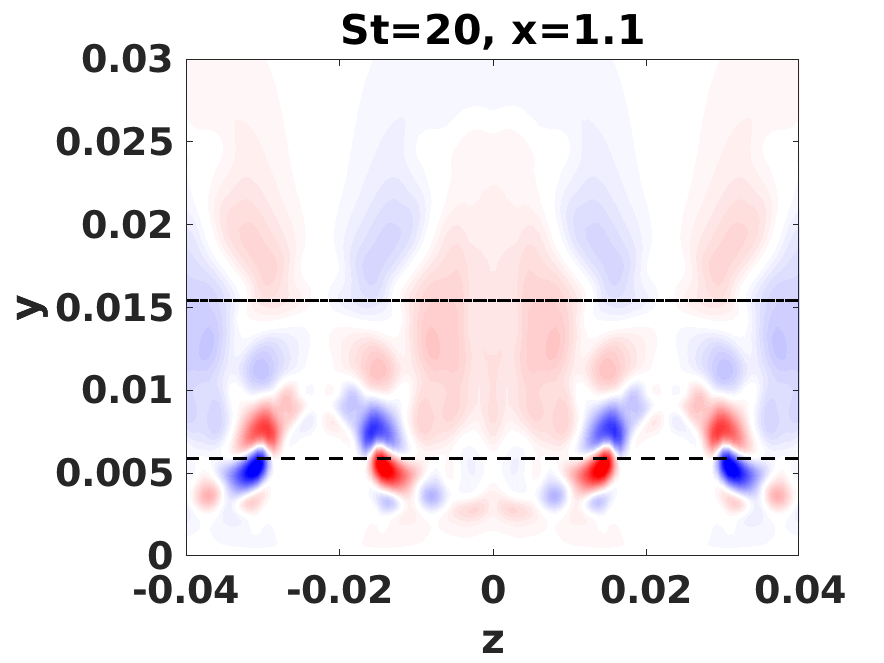}}}
\subfigure[]{{\includegraphics[trim=5 0 18 0,clip,width=0.325\textwidth]{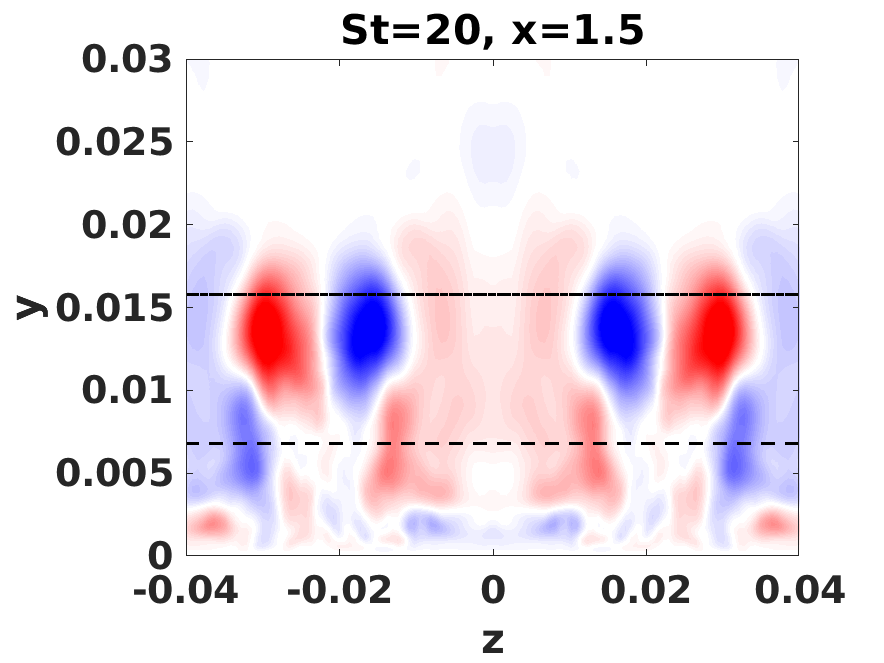}}}
\subfigure[]{{\includegraphics[trim=5 0 18 0,clip,width=0.325\textwidth]{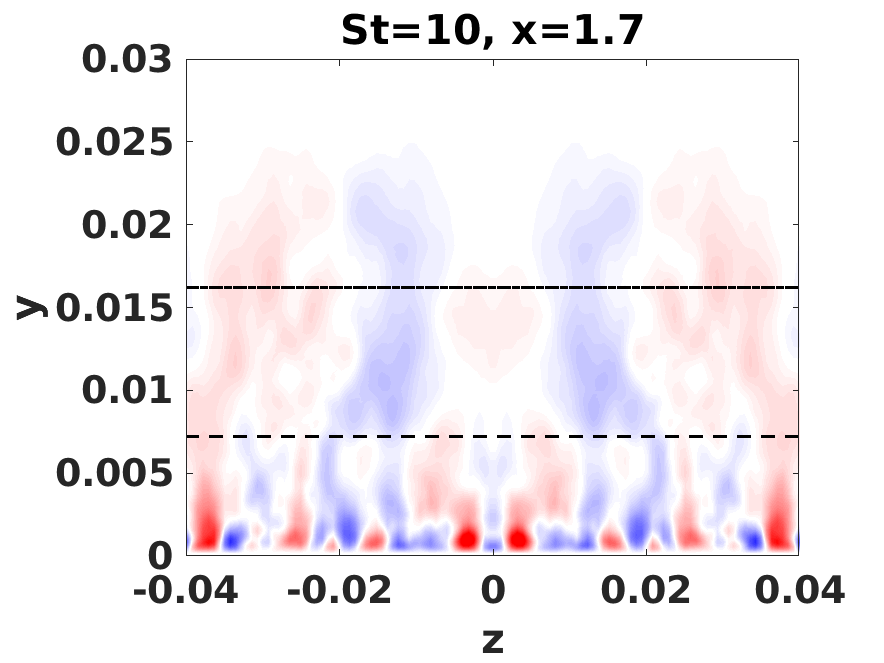}}}
\end{subfigmatrix}
\caption{DMD modes of streamwise velocity fluctuations.
Iso-surfaces at the frequencies $(a)$~$~St~=~20$, and $(b)$~$~St~=~10$, with blue and red representing positive and negative fluctuations, and a magnitude $|u'|=0.0008$.
Corresponding contours in the cross-planes are plotted in $(c,d,e)$ depicting various stages  of transition.}
\label{fig:dmd}
\end{figure}

To gain further insight into the transitional flow, DMD is applied to streamwise velocity perturbations (subtracted from the mean).
The subdomain chosen for the analysis is characterized by spectral broadening in figs.~\ref{fig:Stx},~\ref{fig:up_iso}~($1.1 < x< 1.8$, $0<y<0.03$, $-0.044<z<0.044$).
$800$ uniformly spaced DNS snapshots are chosen at a time step size of~$\Delta t=5\times10^{-3}$ and  the DMD results are summarized in fig.~\ref{fig:dmd}.
Top views of the DMD mode isosurfaces are plotted in figs.~\ref{fig:dmd}(a,b) at~$St~=~20$,~$10$, respectively, with (red)blue indicating (positive)negative fluctuation levels~$|u'|~=~0.0008$.
The corresponding wall-normal signatures are provided in fig.~\ref{fig:dmd}(c,d,e), with each panel saturated to~$80\%$ of the maximum amplitude. 
In fig.~\ref{fig:dmd}(a) the oscillations at the forced frequency~$St=20$ are evident in between positive and negative streaks for~$x<1.2$.
These further intensify around the negative streak in~$1.3< x< 1.45$, spread spanwise in the range~$1.5<x<1.6$, and exhibit weaker signatures for~$x>1.65$.
Alternating positive and negative fluctuations on the negative streak and symmetric fluctuations on the positive streak further suggest the dominance of the subharmonic sinuous oscillations as the secondary instability mechanism.
Paredes et al.~\cite{paredes2020mechanism} also reported this as the dominant streak instability on blunt cones.
Its signature is strongest at the boundary-layer edge~($x=1.1$, fig.~\ref{fig:dmd}(c)) in the initial transitional regime, and moves closer to the entropy-layer edge downstream at~$x=1.5$ (fig.~\ref{fig:dmd}(d)).
Initially weaker low-frequency modes~$(St=10)$ amplify for~$x>1.5$ (fig.~\ref{fig:dmd}(b)), with a predominant near-wall signature~(fig.~\ref{fig:dmd}(e)).
This is an indication of late nonlinear transition stages~\cite{sayadi2014reduced}.
The DMD analysis thus suggests that the transitional features observed in fig.~\ref{fig:up_iso} are attributable to the instability of the primary streak due to high frequency entropic-instabilities, and subsequent breakdown leading to the generation of multiple near-wall streaks. 

The near-wall streamwise vortices and streaks are crucial ingredients for sustaining (near-wall) turbulence. 
The streamwise vortices generate streaks through lift-up, which further undergo secondary-instabilities and breakdown into vortices, creating a self-sustaining process~\cite{schoppa2002coherent}.
Thus, the low-frequency streaks observed in the current study after $x>1.7$, also undergo secondary-instabilities closer to the wall; however, their detailed examination is out of the scope of the current paper.
While some of the streak breakdown features discussed above are observed in incompressible flows~(see, \cite{brandt2002transition}), a distinct feature of hypersonic transition on blunt plates is the amplification of (coupled) thermal fluctuations in the entropy-layer, which are discussed next. 

\subsubsection{Role of temperature perturbations}
\begin{figure}
\centering
\begin{subfigmatrix}{3}
\subfigure[]{{\includegraphics[trim=65 0 75 0,clip,width=0.95\textwidth]{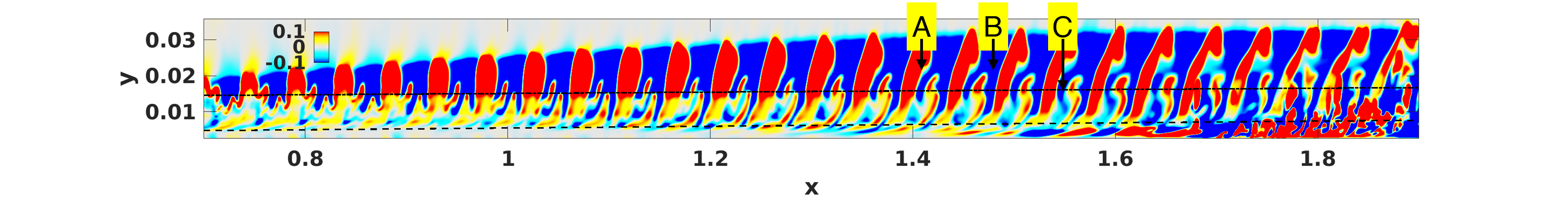}}}
\subfigure[]{{\includegraphics[trim=65 0 75 0,clip,width=0.95\textwidth]{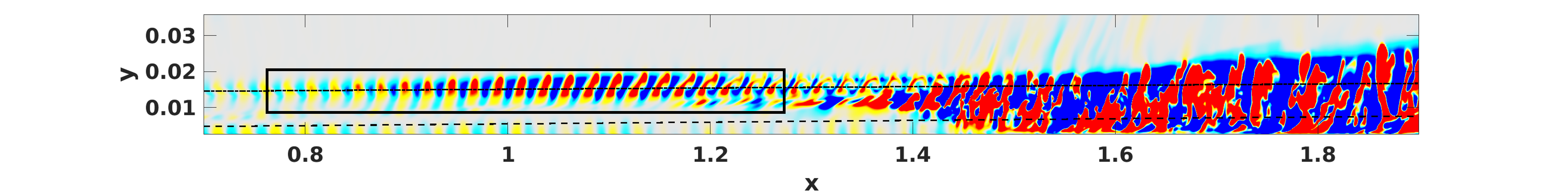}}}
\end{subfigmatrix}
\caption{Contours of instantaneous temperature perturbations illustrating structures in the spanwise planes at the centers of 
$(a)$~high-speed streak ($HS$),~$z\sim0$, $(b)$~low-speed streak ($LS$),~$z\sim-0.02$.
In $(a)$ $A$, $B$, and $C$ denotes various transitional structures and in $(b)$ the rectangle highlights Orr-like mechanism.
Entropy- and boundary layer-edges are marked by dotted-dashed~$(.-)$ and dashed~$(--)$ lines, respectively.
}
\label{fig:Tp}
\end{figure}

The role of temperature perturbations during transition is examined by plotting their (mean subtracted) contours at spanwise locations~$z\sim0$,~$-0.02$, approximately corresponding to the centers of high-($HS$) and low-speed ($LS$) streaks, respectively, in figs.~\ref{fig:Tp}(a,b).
At the symmetry plane~$(HS)$ in fig.~\ref{fig:Tp}(a), above the entropy-layer, large wavelength positive and negative temperature perturbations are observed (structures marked $A$, $B$ respectively), with the former developing to inclined \textit{hook}-shaped structures~$(x\sim1.4)$, with tails extending downward to the boundary-layer.
These structures resemble the inclined positive temperature perturbations reported by Hartman et al.~\cite{hartman2021nonlinear}.
Between these, the negative temperature perturbations are characterized by stretching (wall-normal direction) and tilting (streamwise).
Shorter-wavelength disturbance amplification (marked $C$) at the entropy-layer edge, with spatial support extending to the boundary-layer downstream is also evident.
The signatures of turbulence onset can be discerned for~$x>1.8$, where intense near-wall fluctuations arise.
In this regime, the inclined temperature perturbations above the entropy-layer preserve their structure.

The temperature perturbation contours in the $LS$ plane are depicted in fig.~\ref{fig:Tp}(b).
A (non-modal) Orr-like mechanism is discernible at the entropy-layer edge, where upstream tilted disturbances at~$x\sim 0.7$ propagate by intensifying in the streamwise direction followed by downstream tilting~$(x\sim 1.15)$.
Such a classical Orr mechanism is observed when both the mean shear and mean flow advection are nonzero, and the upstream leaning disturbance tilts downstream~\cite{orr1907stability}.
Here, we observe a similar tilting mechanism, but due to a (larger) mean gradient in the base flow temperature, and hence the term Orr-like mechanism.
Additional fluctuations around~$x\sim 1.3$ are observed above the boundary-layer edge, which is the wall-normal signature of secondary-instabilities discussed in fig.~\ref{fig:urms}(b,c).
Spreading of the disturbances in the wall-normal direction and onset of turbulence is evident for~$x>1.5$.
In the hypersonic transition studies on blunt cones, Hartman et al.~\cite{hartman2021nonlinear} reported that while positive temperature fluctuations form inclined structures and propagate above the entropy-layer, its negative counterparts penetrate the boundary-layer and generate multiple scales leading to turbulence onset.
The current transition scenario is somewhat different, with both positive and negative temperature perturbations amplifying and participating in multiple-scale generation via interactions with the streaks. 

\new{To extract the underlying coherent structures in the temperature perturbation field, DMD is employed on the $HS$ and $LS$ planes.
For this,~$1{,}200$ two-dimensional snapshots separated by a time step~$\Delta t~=~2.5\times 10^{-3}$ are utilized.
Some of the prominent modes are shown in fig.~\ref{fig:New_DMD}, where each panel is saturated to~$80\%$ of the maximum amplitude.
\begin{figure}
\centering
\begin{subfigmatrix}{2}
\subfigure[]{{\includegraphics[trim=120 20 140 0,clip,width=0.95\textwidth]{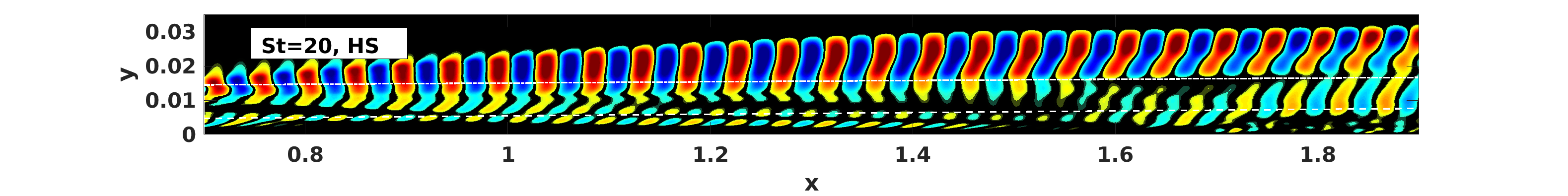}}}
\subfigure[]{{\includegraphics[trim=120 20 140 0,width=0.95\textwidth]{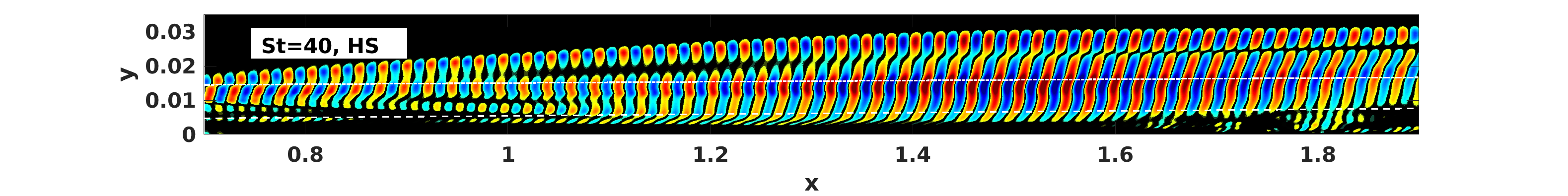}}}
\subfigure[]{{\includegraphics[trim=120 0 140 0,width=0.95\textwidth]{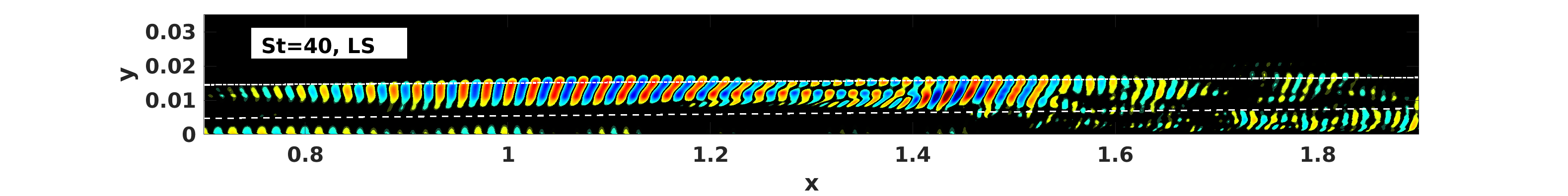}}}
\end{subfigmatrix}
\caption{Temperature modal structures of disturbances in spanwise planes.
In~$(a,~b)$ disturbance structures on high-speed streak~$(HS)$ at frequencies $St=20,~40$ are shown respectively.
In~$(c)$ $St=40$ structure on low-speed streak~$(LS)$ is depicted.
}
\label{fig:New_DMD}
\end{figure}
The temperature mode at $St=20$ on $HS$ plane in fig.~\ref{fig:New_DMD}(a) exhibits significant dynamics around the entropy-layer edge.
The disturbances induced at the forcing location elongate in the wall-normal direction; the lower part moves slower than the upper part because of the velocity gradient and results in the aforementioned streamwise tilting of the inclined hook-shaped structures of fig.~\ref{fig:Tp}(a).
Its harmonic~$(St=40)$ in fig.~\ref{fig:New_DMD}(b) exhibits amplification at the entropy-layer edge coupled with wall-normal stretching extending into the boundary-layer.
The shorter wavelength structures at the entropy-layer edge in fig.~\ref{fig:Tp}(a) correspond to these modes.
The signatures of translating structures above the entropy-layer edge are also evident, and may be related to the form of the forcing function.
The mode at~$St=40$ on $LS$ plane in fig.~\ref{fig:New_DMD}(c) is amplified by the Orr-like mechanism in $0.8<x<1.2$ in the entropy-layer.
At~$x\sim1.4$, structures observed at~$y\sim 0.01$ correspond to the wall-normal spreading of the disturbances highlighted in fig.~\ref{fig:Tp}(b).}

\begin{figure}
\centering
\begin{subfigmatrix}{3}
\subfigure[]{{\includegraphics[trim=80 0 100 0,clip,width=0.325\textwidth]{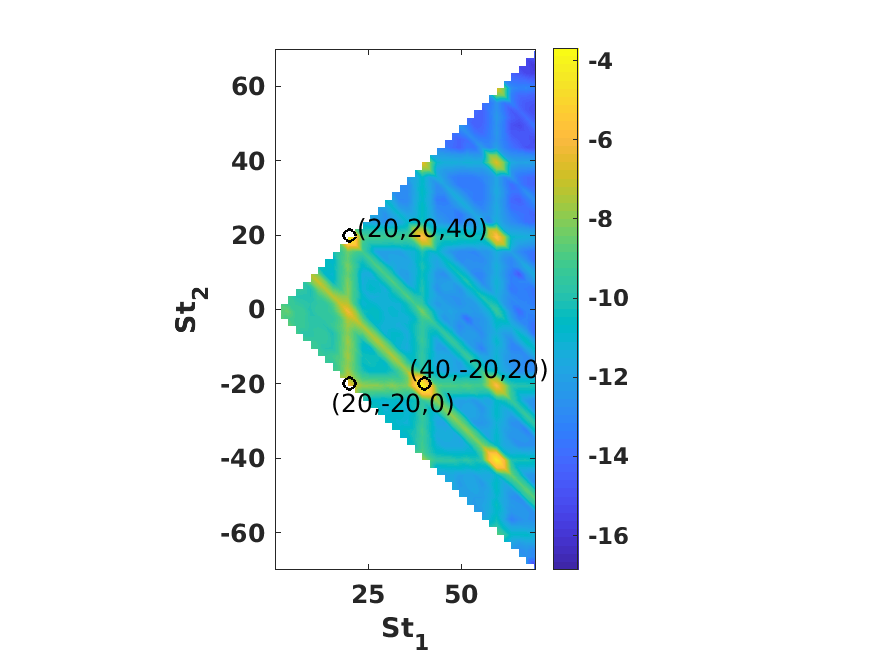}}}
\subfigure[]{{\includegraphics[trim=80 0 100 0,clip,width=0.325\textwidth]{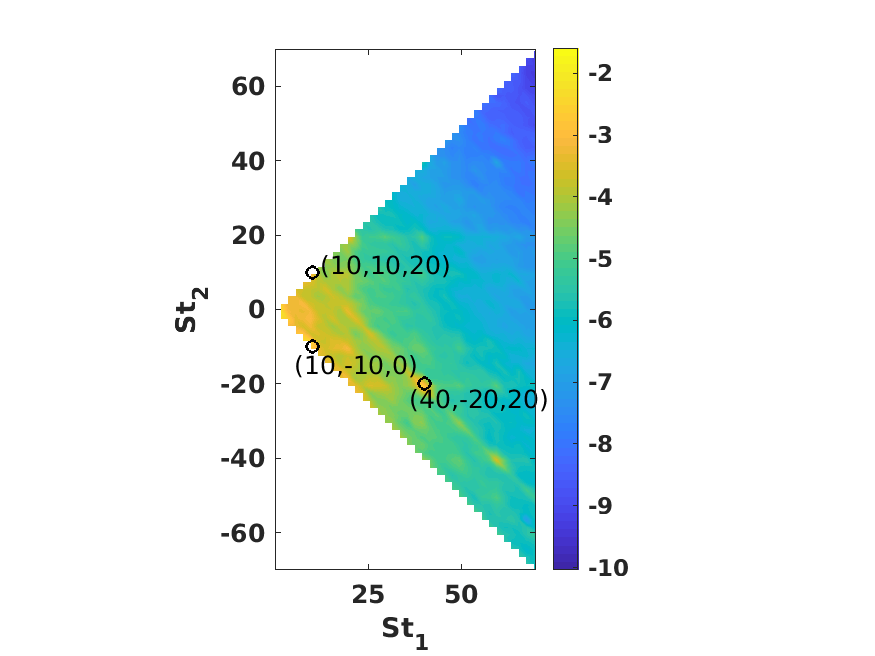}}}
\subfigure[]{{\includegraphics[trim=80 0 100 0,clip,width=0.325\textwidth]{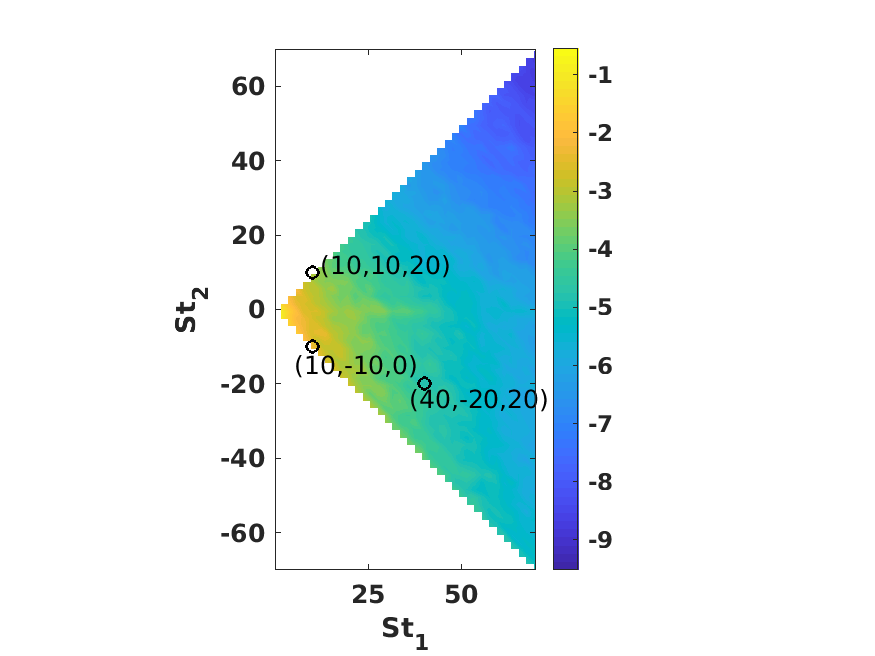}}}
\end{subfigmatrix}
\caption{Magnitude (logarithmic) of temperature-streamwise velocity mode cross-bispectrum in cross-flow planes at~$(a)$~$x=1.1$, $(b)$~$x=1.5$, and $(c)$~$x=1.9$.}
\label{fig:cmd}
\end{figure}

DMD extracts the mode shape at a chosen frequency and CBMD provides insight into the potential interactions leading to these structures.
The triadic interactions between temperature perturbations at the frequencies~$St_1$ and~$St_2$ resulting in the velocity perturbations at~$St_3=St_1+St_2$ due to phase coupling~$(\phi_3=\phi_1+\phi_2)$ are considered.
\new{The triple phase locking is also accompanied by wavenumber interactions satisfying resonance conditions and are not shown here.}
For CBMD analysis,~$1600$ snapshots spaced~$\Delta t~=~2.5\times10^{-3}$ apart are divided into~$50\%$ overlapped blocks with~$200$ elements each.
The streamwise locations, $x=1.1$,~$1.5$,~$1.9$ are chosen to highlight the various transition stages.
The corresponding mode bispectrum, which represents the strength of triadic interactions, is plotted in fig.~\ref{fig:cmd}.
The spectral features are symmetric in the other quadrants and hence not repeated.
The mode bispectrum at~$x=1.1$ in fig.~\ref{fig:cmd}(a) exhibits a typical grid-like pattern with maxima at its nodes.
These local maxima are the signature of a cascade of triads generated through interactions, such as the oblique breakdown signature-modes discussed in fig.~\ref{fig:Stx}(b).
The bands (diagonal, horizontal, vertical) are the a result of spectral leakage, a phenomenon inherent to discrete Fourier transform of nonperiodic data.
Here, the strongest interactions are observed between temperature fluctuations at $St=40$ and $20$ intensifying velocity perturbations at $St=20$~$(\hat{T}_{40}-\hat{T}_{20}\rightarrow\hat{u}_{20})$, indicating an energy transfer mechanism enhancing the streak secondary instabilities.
Other interactions leading to higher harmonic generation of forcing waves at $St=20$ in velocity perturbations, such as $(\hat{T}_{20}+\hat{T}_{20}\rightarrow\hat{u}_{40}),(\hat{T}_{40}+\hat{T}_{20}\rightarrow\hat{u}_{60})$, are also apparent.
Further downstream at~$x=1.5$ (fig.~\ref{fig:cmd}(b)), while the dominant interaction $(\hat{T}_{40}-\hat{T}_{20}\rightarrow\hat{u}_{20})$ persists, the typical grid pattern fades, highlighting the predominance of low-frequency interactions.
In fig.~\ref{fig:cmd}(c), the dominance of low-frequency interactions is more pronounced.
These inferences are consistent with the velocity DMD structures reported in fig.~\ref{fig:dmd}, where low-frequency modes dominate in the late transitional regimes.

\begin{figure}
\centering
\begin{subfigmatrix}{3}
\subfigure[]{{\includegraphics[trim=5 0 18 0,clip,width=0.325\textwidth]{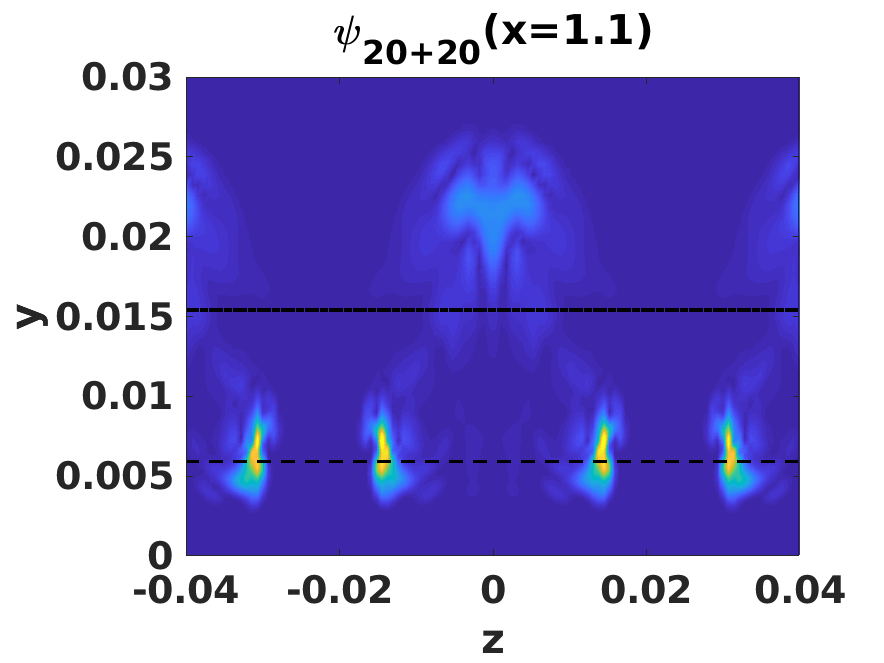}}}
\subfigure[]{{\includegraphics[trim=5 0 18 0,clip,width=0.325\textwidth]{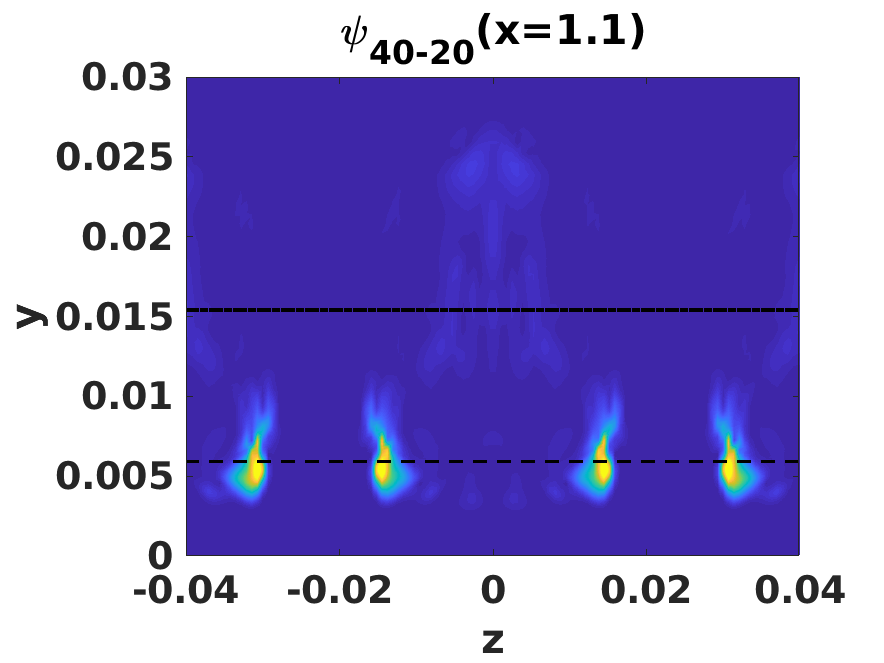}}}
\subfigure[]{{\includegraphics[trim=5 0 18 0,clip,width=0.325\textwidth]{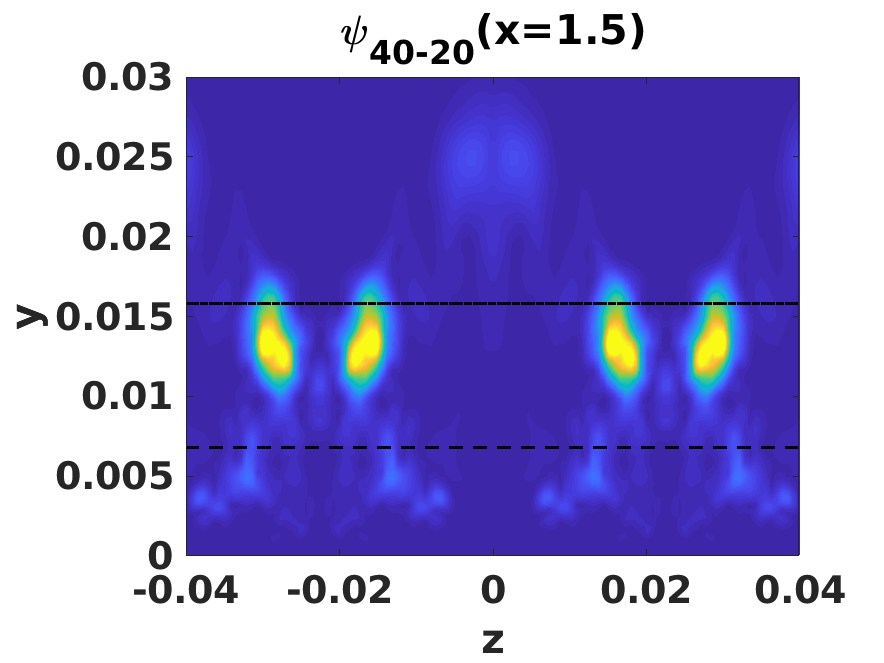}}}
\subfigure[]{{\includegraphics[trim=5 0 18 0,clip,width=0.325\textwidth]{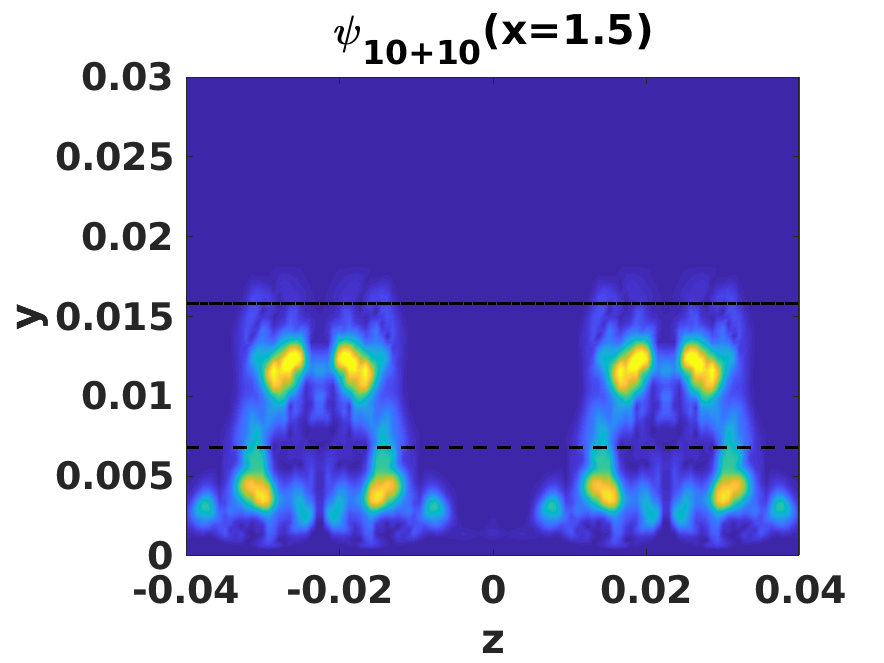}}}
\subfigure[]{{\includegraphics[trim=5 0 18 0,clip,width=0.325\textwidth]{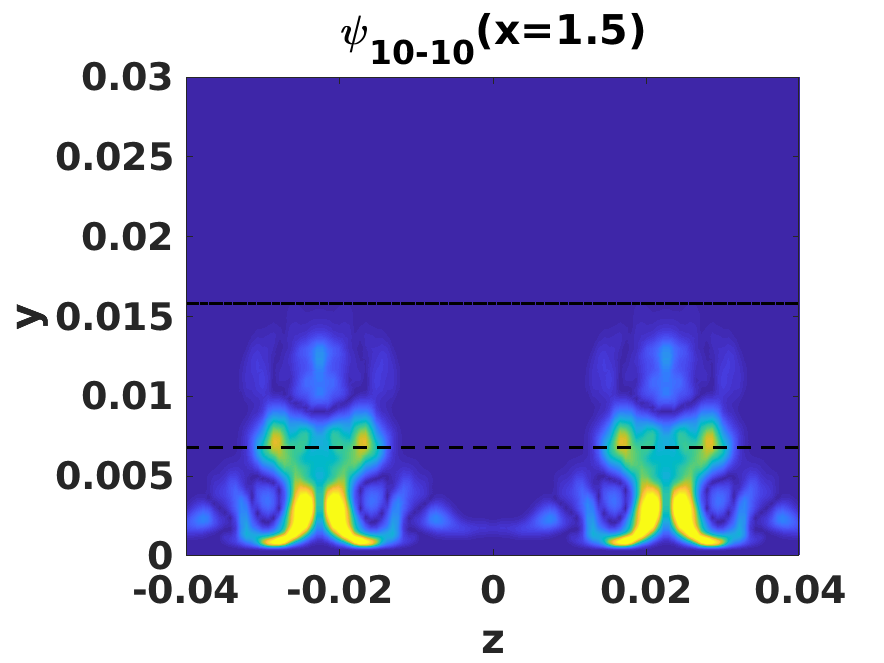}}}
\subfigure[]{{\includegraphics[trim=5 0 18 0,clip,width=0.325\textwidth]{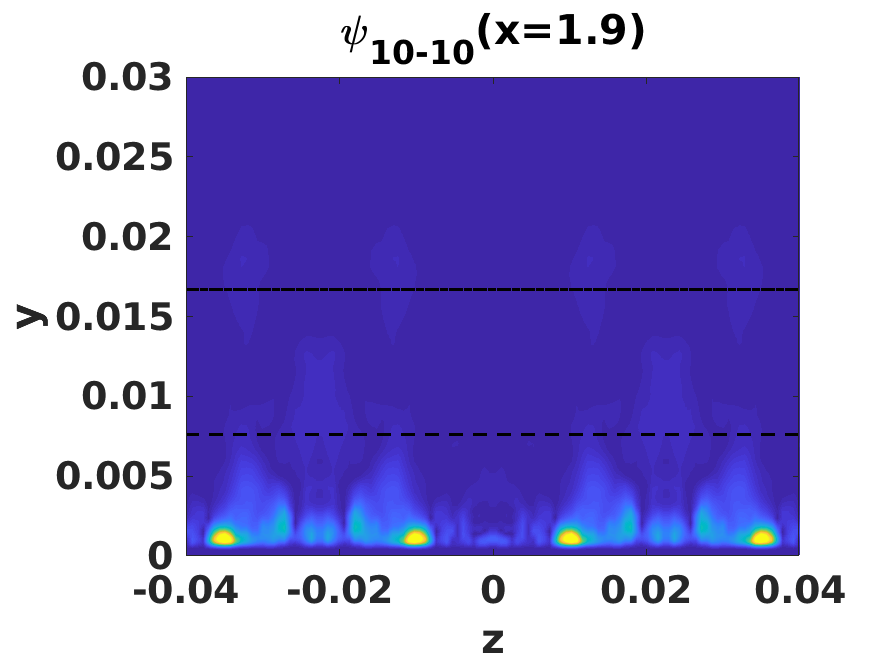}}}
\end{subfigmatrix}
\caption{Cross-bispectral modal interaction maps~($\psi_{St_1\pm St_2}$) in the crossflow planes highlighting phase coupling between temperature and velocity fluctuations. 
}
\label{fig:cmd_mode}
\end{figure}

Nonlinear transfer of energy among the unstable modes, and to the mean flow can transpire at various wall-normal locations based on their phase velocities.
Bountin et al.~\cite{bountin2008evolution} experimentally investigated nonlinear processes during hypersonic transition across the boundary-layer using bicoherence analysis.
They reported second-mode related oscillations at the maximum rms voltage fluctuation layer, and nonlinear interactions of higher intensity above and below it.
To quantify the triadic interactions in the current work, interaction maps~($\psi$) of the modes are examined in fig.~\ref{fig:cmd_mode}, where the transfer of energy from the temperature perturbations into velocity fluctuations are highlighted, and each panel is saturated with~$80\%$ of the maximum value.
The interaction map is defined as the entry-wise product of the bispectral mode~$(\hat{u}_{St_1+St_2})$ and the cross-frequency mode~$(\hat{T}_{St_1\circ St_2})$.
In figs.~\ref{fig:cmd_mode}(a,b), interactions of temperature fluctuations at the forcing frequency~$St=20$, marked in fig.~\ref{fig:cmd}(a), are plotted at~$x=1.1$.
The self sum-interactions of temperature fluctuations enhances the velocity signature on the flanks of low-speed streak~$(\hat{T}_{20}+\hat{T}_{20}\rightarrow\hat{u}_{40})$ at the boundary-layer edge and above the entropy-layer in fig.~\ref{fig:cmd_mode}(a).
The difference-interaction of harmonic and forcing temperature perturbations~$(\hat{T}_{40}-\hat{T}_{20}\rightarrow\hat{u}_{20})$ contributing to the modal structure in fig.~\ref{fig:dmd}(c) is plotted in fig.~\ref{fig:cmd_mode}(b).
This interaction has a similar spatial distribution as the sum-interaction, but is weaker above the entropy-layer.

Figure~\ref{fig:cmd_mode}(c,d) depicts two distinct mechanisms of $St=20$ velocity mode generation corresponding to fig.~\ref{fig:dmd}(d).
The high-frequency interactions~$\hat{T}_{40}-\hat{T}_{20}\rightarrow\hat{u}_{20}$ predominantly occur on the crests of the low-speed streaks in the entropy-layer and exhibit a weaker signature above it.
The wall-normal shift of this interaction in fig.~\ref{fig:cmd_mode}(b,c) is reflective of the rms velocity signature in fig.~\ref{fig:urms}(a,d).
However, the low-frequency modes~~$\hat{T}_{10}+\hat{T}_{10}\rightarrow\hat{u}_{20}$ interact in entropy- and boundary layers.
Self-difference interactions at low-frequency~$St=10$ distort the mean flow (fig.~\ref{fig:cmd_mode}(e)), which partially explains the peak magnitude of near-wall rms fluctuations at the streak valleys in fig.~\ref{fig:urms}(d).
This highlights the role of low-frequency waves in transferring disturbance energy from the entropy-layer into near wall regions, aiding in boundary-layer transition.
These interactions at~$x=1.9$ in fig.~\ref{fig:cmd_mode}(f) are now dominant near wall, consistent with the spanwise homogenization in fig.~\ref{fig:urms}(f).
The progression of transition also reveals spanwise wavenumber filling and can be attributed to the generation of newer near-wall modal/non-modal instabilities arising due to mean flow distortion.

\begin{figure}
\centering
\begin{subfigmatrix}{1}
\subfigure[]{{\includegraphics[trim=120 20 140 0,clip,width=0.95\textwidth]{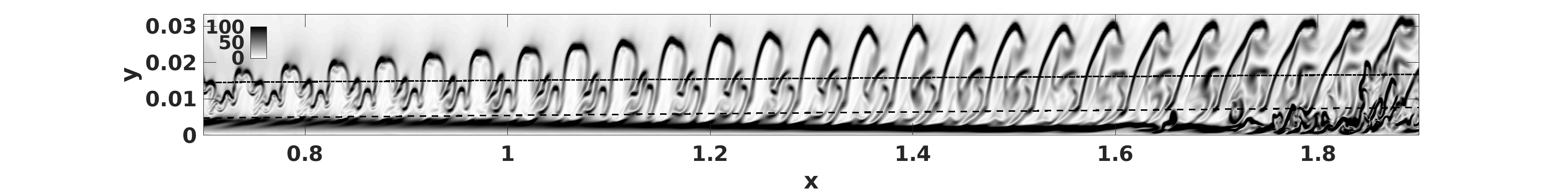}}}
\subfigure[]{{\includegraphics[trim=120 20 140 0,clip,width=0.95\textwidth]{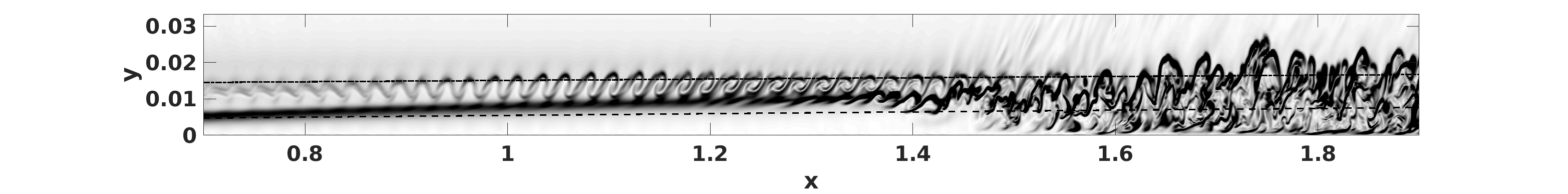}}}
\end{subfigmatrix}
\caption{Contours of density-gradient magnitude (pseudo-Schlieren) at the centers of $(a)$~high-speed streak ($HS$), $(b)$~low-speed streak ($LS$).
}
\label{fig:grad_rho}
\end{figure}

For qualitative comparisons with experiments on transitional flows around blunt bodies, the contours of density gradient magnitude (pseudo-Schlieren) are plotted in the spanwise planes $HS$, $LS$ in figs.~\ref{fig:grad_rho}(a,b).
The signature of the high-speed streak development is observed near the wall as a decrease in the streak wall-normal thickness relative to the laminar boundary-layer in fig.~\ref{fig:grad_rho}(a) till~$x\sim 1.6$.
Disturbances amplifying in the entropy-layer elongate in the wall-normal direction and evolve into inclined structures, consistent with the experiments of Kennedy et al.~\cite{kennedy2019visualizations} and Hill et al.~\cite{hill2021experimental} on blunt cones and ogive-cylinders, respectively.
These disturbances penetrate the streak, followed by destabilization and breakdown for~$x>1.65$.  
A lift-up in the density-gradient signature associated with the low-speed streak formation is evident in fig.~\ref{fig:grad_rho}(b).
The entropy-layer disturbances amplify through the Orr-like mechanism in~$0.8<x<1.1$ and modulates the streak in~$1.2<x<1.3$, resulting in multiple-scale  generation above the (laminar) boundary-layer edge.
This qualitatively validates the inference that the disturbances developing on the crest in fig.~\ref{fig:urms} are induced by the perturbations in the entropy-layer and are outer instabilities according to the classification of~\citet{vaughan2011stability}.
A notable distinction is that the outer instabilities in the low Mach number regime can exist only in the presence of streaks; this condition is not necessary, however, when an entropy-layer is present. 
In this case, these disturbances spread further in the wall-normal direction and contaminate the otherwise quiet boundary-layer as a precursor to the onset of turbulence.
While the structures of entropy-layer disturbances are preserved in the late nonlinear stages in high-speed streak regions (fig.~\ref{fig:grad_rho}(a)), they breakdown in the low-speed streak region.

\subsubsection{Development of turbulent flow}
\begin{figure}
\centering
\begin{subfigmatrix}{3}
\subfigure[]{{\includegraphics[trim=1 85 25 90,clip,width=0.7\textwidth]{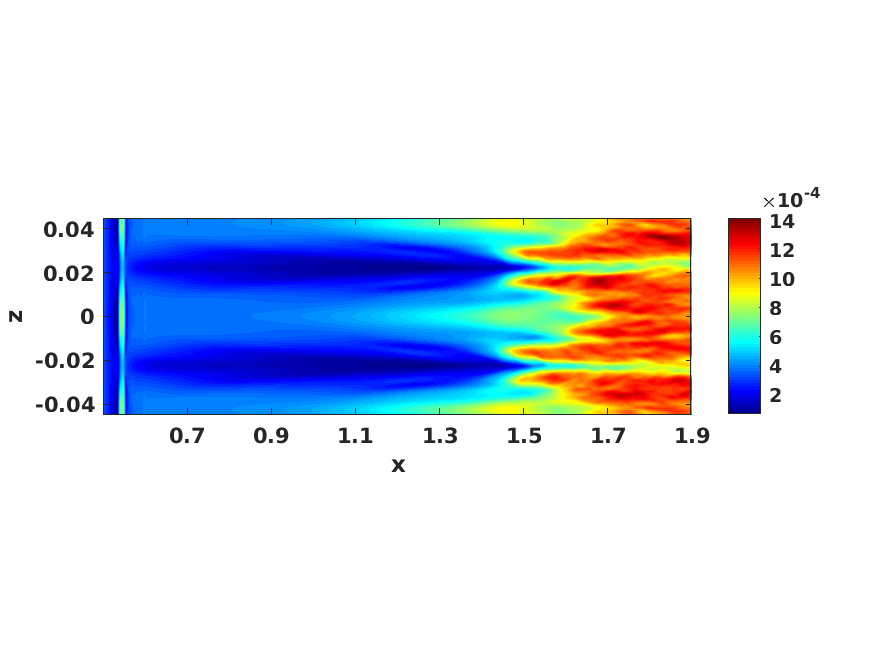}}}
\subfigure[]{{\includegraphics[trim=5 0 18 0,clip,width=.325\textwidth]{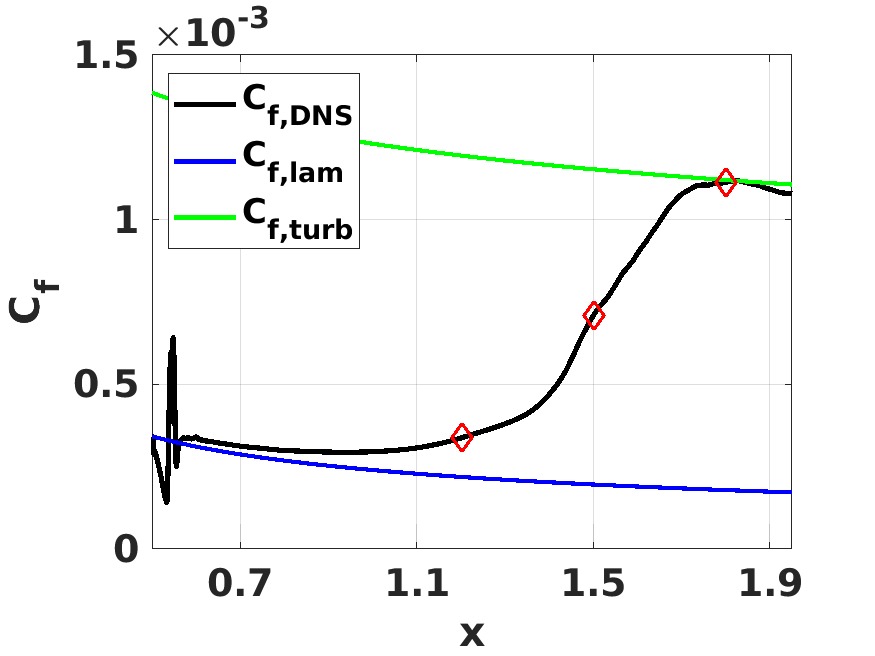}}}
\subfigure[]{{\includegraphics[trim=5 0 18 0,clip,width=.325\textwidth]{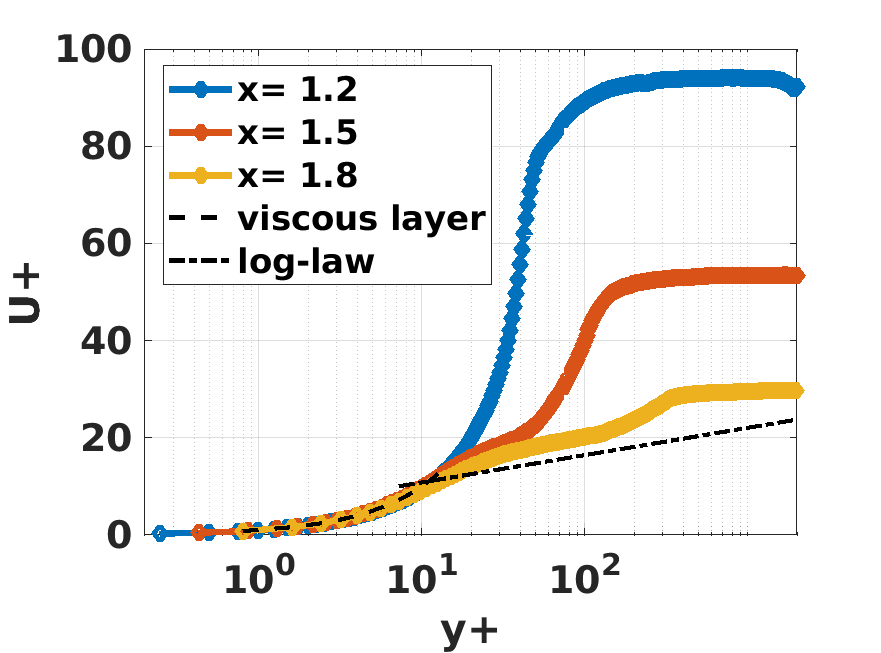}}}
\end{subfigmatrix}
\caption{Flow features highlighting onset of turbulence.
$(a)$~Contours of time-averaged skin friction coefficient~$(C_f)$ in the streamwise direction~$(x)$.
$(b)$~Span- and time-averaged skin-friction coefficient, compared with empirical laminar and turbulent correlations.
$(c)$~Streamwise development of equilibrium turbulent velocity profile through van Driest transformation at the center of low-speed streak~$z\sim-0.02$.
}
\label{fig:Cf}
\end{figure}

Skin friction and van Driest~\cite{van1956problem} transformed velocity profiles are examined to assess the progression of transition towards fully turbulent flow. 
In fig.~\ref{fig:Cf}(a,b) the skin friction coefficient~$(C_{f})$ over the flat plate is plotted, and is defined in eq.~\ref{eqn:cf}.
\begin{equation}
\label{eqn:cf}
    C_f=\frac{2}{Re}\mu \frac{\partial u}{\partial y}|_{y=0}
\end{equation}
Contours of the time-averaged skin friction coefficient are displayed in fig.~\ref{fig:Cf}(a).
Dominant signatures of streamwise streaks can be observed from~$x \sim 0.6$ to $x \sim 1.4$ around~$z \sim \pm 0.02$.
This is the wall trace of nonlinearly generated streaks discussed in the context of fig.~\ref{fig:up_iso}.
At~$x \sim 1.5$, large skin friction values are observed at~$z \sim 0.015$,~$0.025$, suggesting the transition of the low-speed streaks.    
Around~$z \sim 0$, the skin friction coefficient increases gradually, consistent with  positive streak growth.
Here, large values of skin friction can be observed around~$x \sim 1.6$, confirming that the positive streak transitions later than its negative counterpart.
After $x\sim 1.7$, streaks merge in the spanwise direction suggesting late nonlinear stages of transition.
The contours in fig.~\ref{fig:Cf}(a) are spanwise averaged and plotted in fig.~\ref{fig:Cf}(b) along with empirical laminar and turbulent Eckert correlations~\cite{eckert2003engineering}.
In~$0.6 < x < 1.3$, the skin friction coefficient increases gradually, followed by a rapid deviation towards turbulent value in the range~$ 1.4 < x< 1.7$.
The latter extent corresponds to the streak breakdown regime in fig.~\ref{fig:Cf}(a).
Further downstream, a decrease in friction coefficient is observed, indicating that the flow is settling to an equilibrium turbulence.
The mean skin friction in this regime is about~$7$ times the corresponding laminar coefficient. 
\new{For other low-amplitude forcing cases discussed in table~\ref{tab:sim_cond}, no significant deviation from laminar friction correlation was found.}

The time-averaged development of streamwise velocities at the center of low-speed streaks~$(z \sim -0.2)$ are plotted in fig.~\ref{fig:Cf}(c) using scaled wall-units,~$u^{+}$ and $y^{+}$.
The compressibility effects are accounted through the van Driest transformation.
The viscous and log-layers are depicted by dashed and dashed-dotted lines, respectively.
While a linear relation between~$u^{+}$ and $y^{+}$ is characteristic of the viscous sublayer, the presence of the log-layer represents a fully turbulent flow. 
The viscous sublayer extends up to~$y^{+} \sim 10$ at $x=1.8$, consistent with earlier hypersonic transition studies~\cite{hartman2021nonlinear, hader2019direct}.
The progression of flow from $x=1.2$ to $x=1.8$ indicates the onset of early turbulent features.

\new{The entropy-layer oblique breakdown mechanism discussed 
highlights the dominance of streaks in flow destabilization.
This underlying 
dynamics 
suggests that weakening of streaks could delay transition for control applications.
In supersonic boundary layers, Sharma et al.~\cite{sharma2019control} showed that actuating modes with four to five times the fundamental streak wavenumber were beneficial in controlling streak-induced transition. 
They also reported increased gain by introducing additional downstream control strips.
These techniques could be potentially explored for transition control on blunted plates.}

\section{Summary}
\label{sec:summary}
Transition to turbulence on a blunted flat plate at Mach~$4$ has been examined with high-fidelity numerical techniques.
The freestream conditions were chosen based on the available experimental conditions, with a nose radius in the \textit{medium} bluntness regime.
Linear stability investigations revealed modal instabilities in the boundary- and entropy-layers; however, their growth-rates are too low to initiate transition.
To examine the receptivity and linear instability of two-dimensional perturbations, wavetrains were triggered at various wall-normal locations.
Among these, perturbations triggered in the entropy-layer exhibited the highest receptivity, with amplification of well-defined coherent structures.

The three-dimensional evolution of a wavepacket triggered in the entropy-layer, and the corresponding wall pressure signatures revealed growing low-frequency oblique first-modes, planar and oblique entropic-instabilities.
Complimentary wavepacket simulations performed by forcing at the boundary-layer edge and the wall indicate that while the first-mode waves are consistently generated in the cases considered, the signature of the entropic-instabilities on the wall is well pronounced when the forcing is applied in the entropy-layer.
Boundary and entropic layer instabilities develop symmetrically in the spanwise direction, and the oblique breakdown of the latter is explored as a potential route to turbulence.

The evolution of high-amplitude oblique waves triggered in the entropy-layer was examined with Q-criterion isosurfaces, which reveal zigzag shaped vortical structures above the streamwise streaks; these intensify and breakdown leading to finer-scale structures.
The zigzag vortices are predominantly located above the low-speed streaks, which transitions earlier than its high-speed counterpart.
Dynamic mode decomposition of the streamwise velocity perturbations predicted symmetric and antisymmetric oscillations for high- and low-speed streaks respectively, at the forcing frequency, suggesting the sinuous subharmonic oscillations as the dominant streak secondary instability mechanism.  

The role of temperature perturbations during transition has also been investigated. 
The signatures of both positive and negative temperature fluctuations are apparent throughout the transition process. 
Orr-like mechanisms at distinct frequencies with spatial support in the entropy layer are observed in different planes.  
The triadic interactions characterizing the transfer of energy from temperature to velocity perturbations have been examined with cross-bispectral mode decomposition. 
High-frequency temperature perturbations are predominant on the flanks of the low-speed streaks at the laminar boundary-layer edge in the initial transition regime. 
This zone moves to the crest of the low-speed streaks in the entropy-layer downstream, and is followed by rapid amplification leading to the generation of perturbations at multiple frequencies.
These disturbances further distort the mean flow through self-difference interactions and transfer the perturbation energy from the entropy-layer into the boundary-layer.
Owing to the mean flow modification, additional instabilities can develop in the boundary-layer, leading to its breakdown.
These late stages of the transition are dominated by low-frequency streamwise streaks resulting in a skin friction coefficient~$7$ times the laminar value.

\new{
The main similarity with medium bluntness cones is observed in  boundary layer streak formation and its destabilization by entropy layer disturbances. 
Key differences include the absence of destabilizing role of negative temperature fluctuations in the symmetry plane and a second reduction in signature-mode amplitudes before breakdown.
These could be potentially explained by body divergence and shorter entropy-layer swallowing length on blunted cones.}

\new{Future work includes examination of alternate transition scenarios such as the interaction of entropic and boundary layer instabilities.
Also, modeling natural receptivity through wind tunnel or atmospheric flight informed freestream disturbances could further improve transition prediction.
This will aid in better transition control strategies.}

\section*{Acknowledgements}
This research was supported by the Office of Naval Research (Grant: N00014-21-1-2408) monitored by Dr. E. Marineau with R. Burnes as the technical point of contact.
Discussions with Dr. M. Choudhari, Dr. P. Paredes, Prof. T. Zaki, and Prof. S. Unnikrishnan are gratefully acknowledged.
The opinions, findings, views, conclusions, or recommendations contained herein are those of the authors and should not be interpreted as necessarily representing the official policies or endorsements, either expressed or implied, of ONR or the U.S. Government.
The simulations were carried out using resources provided by the U.S. Department of Defense High Performance Computing Modernization Program and the Ohio Supercomputer Center. 
Several figures were made using FieldView software with licenses obtained from the Intelligent Light University Partnership Program.

\bibliography{bibtex}

\end{document}